\documentclass[aps,prl,twocolumn,nofootinbib,groupedaddress,superscriptaddress
,longbibliography]{revtex4-2}

\usepackage{amsmath,amssymb}
\usepackage{graphicx}
\usepackage{multirow}
\usepackage{bm}
\usepackage{mathtools}
\usepackage{dsfont}
\usepackage{amsfonts}
\usepackage{xcolor}
\usepackage[normalem]{ulem}
\usepackage{url}
\usepackage{wasysym}
\usepackage{bbm}
\usepackage{hyperref}
\usepackage{braket}
\usepackage{booktabs}

\begin{document}

\title{Deciphering Chiral Superconductivity via Impurity Bound States}

\author{Yuchang Cai}
\address{Department of Physics and Astronomy, The University of Tennessee, Knoxville, Tennessee 37996, USA}

\author{Rui-Xing Zhang}
\email{ruixing@utk.edu}
\address{Department of Physics and Astronomy, The University of Tennessee, Knoxville, Tennessee 37996, USA}
\address{Department of Materials Science and Engineering, The University of Tennessee, Knoxville, Tennessee 37996, USA}

\begin{abstract}
Determining the symmetry of Cooper pairs remains a central challenge in the study of unconventional superconductors, particularly for chiral states that spontaneously break time-reversal symmetry. Here we demonstrate that point-like impurities in chiral superconductors generate in-gap bound states with a distinctive asymmetry: the local density of states at the impurity site vanishes at one bound-state energy, but not at its particle-hole conjugate. We prove this behavior analytically in generic two-dimensional, single-band chiral superconductors, showing it arises from a fundamental interplay between pairing chirality and crystalline rotation symmetry. Our numerical simulations confirm that this diagnostic feature persists in multiband systems and for spatially extended impurities. Our results establish a symmetry-enforced real-space diagnostic for chiral superconductivity at the atomic scale.
\end{abstract}

\maketitle

{\it Introduction} - Topological superconductivity was first theoretically conceived through the study of chiral superconductors~\cite{read2000paired,ivanov2001nonabelian,sau2010generic,qi2010chiral,kallin2016chiral,can2021high}, in which the Cooper pairing acquires a quantized phase winding around the Fermi surface and spontaneously breaks time-reversal symmetry. These chiral superconductors host Majorana quasiparticles localized at edges and vortex cores~\cite{volovik1999fermion}, making them foundational platforms for topological quantum computation~\cite{nayak2008RMP}. Over the past two decades, extensive experimental efforts have sought to realize chiral superconductivity in materials such as Sr$_2$RuO$_4$~\cite{luke1998time,mackenzie2003RMP}, UTe$_2$~\cite{ran2019nearly,jiao2020chiral}, UPt$_3$~\cite{schemm2014observation,avers2020broken}, and more recently in van der Waals platforms~\cite{ribak2020chiral,zhao2023time,wan2024unconventional,han2025signatures}. Despite numerous tantalizing hints, a smoking-gun signature remains elusive, largely due to the complex and often subtle nature of the phenomena and the difficulty of uniquely identifying chiral pairing in experiments~\cite{pustogow2019constraints,kayyalha2020absence}. This challenge highlights the urgent need for a simple, definitive, and broadly applicable diagnostic to reveal the chiral nature of Cooper pairs.

In this work, we show that chiral superconductivity leaves a robust fingerprint in the spatial structure of impurity-bound states. We focus on point-like defects, such as atomic vacancies or substitutions, that trap in-gap bound states appearing in particle–hole symmetric pairs~\cite{balatsky2006RMP,lothman2014defects,pouyan2024YSR}. Remarkably, we find that the local density of states (LDOS) at the impurity site necessarily vanishes at {\it either} positive {\it or} negative bound-state energy, but not both. This striking nodal behavior is enforced by a compatibility condition between the phase winding of chiral Cooper pairs and crystalline rotation symmetry, an effect generally absent in non-chiral systems. We prove this result analytically for general single-band systems and confirm its robustness through extensive numerical simulations across a variety of models, including those with extended impurities and multiband pairing. Our findings establish a simple and experimentally accessible criterion for identifying chiral superconductivity via real-space detection of impurity states. This theoretical framework also provides a solid basis for interpreting recent scanning tunneling microscopy (STM) results on Sn/Si(111), where nodal impurity-state structures consistent with chiral $d$-wave pairing have been reported~\cite{wu2025Sn}.

\begin{table}[t]
\centering
\renewcommand\arraystretch{1.5} 
\setlength{\tabcolsep}{1.8mm}{
\begin{tabular}{c|cccccc} 
 & $j_z=0$ & $j_z=1$ & $j_z=2$ & $j_z=3$\\
\hline
$C_2$ & $i\sigma_y,k_\pm \sigma_{0,z}$ & $k_\pm\sigma_x$ & N/A & N/A \\
\hline
$C_3$ &   $i\sigma_y$  &  $k_+\sigma_x,k^2_-\sigma_y$   &  $k_+^2\sigma_y$   &   N/A   \\
\hline
$C_4$ &  $i\sigma_y$   &  $k_+\sigma_x,k^3_-\sigma_x$   &  $k_\pm^2\sigma_y$   &   $k_+^3\sigma_x,k_-\sigma_x$   \\
\hline
$C_6$ &  $i\sigma_y$   &   $k_+\sigma_x$  &  $k_+^2\sigma_y$   &   $k_\pm^3\sigma_x$   \\
\hline
\end{tabular}}
\caption{Classification of chiral pairing function in the continuum limit based on $j_z$, i.e., irreducible representation under $C_n$. $i\sigma_y$ denotes the non-chiral $s$-wave spin-singlet pairing.}
\label{table:pairing class}
\end{table}

{\it Chiral Pairings \& Impurity States} - We consider a two-dimensional (2d) metallic system with a pair of spin-degenerate Fermi surfaces that develop Cooper-pairing instability. The low-energy bands forming the Fermi surfaces are denoted by $E_{{\bf k}}$. We assume the system to respect an out-of-plane $n$-fold rotational symmetry $C_n$ ($n = 2, 3, 4, 6$), under which each Fermi surface is invariant. Superconductivity is described within the mean-field formalism by the Hamiltonian ${\cal H} = \sum_{\bf k} \Psi_{\bf k}^\dagger H({\bf k}) \Psi_{\bf k}$ under the Nambu basis $\Psi_{\bf k} = (c_{{\bf k},\uparrow}, c_{{\bf k},\downarrow}, c^\dagger_{-{\bf k},\uparrow}, c^\dagger_{-{\bf k},\downarrow})^T$. The Bogoliubov–de Gennes (BdG) Hamiltonian matrix is,
\begin{equation}
H({\bf k}) = \begin{pmatrix}
    \xi_{{\bf k}} & \Delta({\bf k}) \\
    \Delta^\dagger({\bf k}) & -\xi_{-{\bf k}}
\end{pmatrix},
\label{eq:BdG}
\end{equation}
where $\xi_{{\bf k}} = E_{{\bf k}} - \mu$ and $\mu$ is the chemical potential. The pairing potential is generally decomposed as $\Delta({\bf k}) = d_0({\bf k}) \sigma_0 + {\bf d}({\bf k}) \cdot {\bm \sigma}$, where ${\bm \sigma} = (\sigma_x, \sigma_y, \sigma_z)$ are Pauli matrices in spin space. Fermionic antisymmetry imposes $d_{\alpha}({\bf k}) = -d_{\alpha}(-{\bf k})$ for $\alpha \in \{0,x,z\}$ (spin-triplet), while $d_y({\bf k})$ (spin-singlet) must be even under ${\bf k} \to -{\bf k}$. Pairing channels can be further classified by the irreducible representations (irreps) of the rotation group $C_n$, labeled by an angular momentum index $j_z \in \mathbb{Z}$ defined modulo $n$. In this work, we focus on complex chiral pairing states that (i) exhibit a phase winding of $e^{il\theta}$ ($l\in\mathbb{Z}$) around the Fermi surface and (ii) generate a full BdG gap. In Table. \ref{table:pairing class}, we summarize and classify such chiral orders in the continuum limit according to their $C_n$ irreps.

\begin{figure*}[t]
    \includegraphics[width=0.8\textwidth]{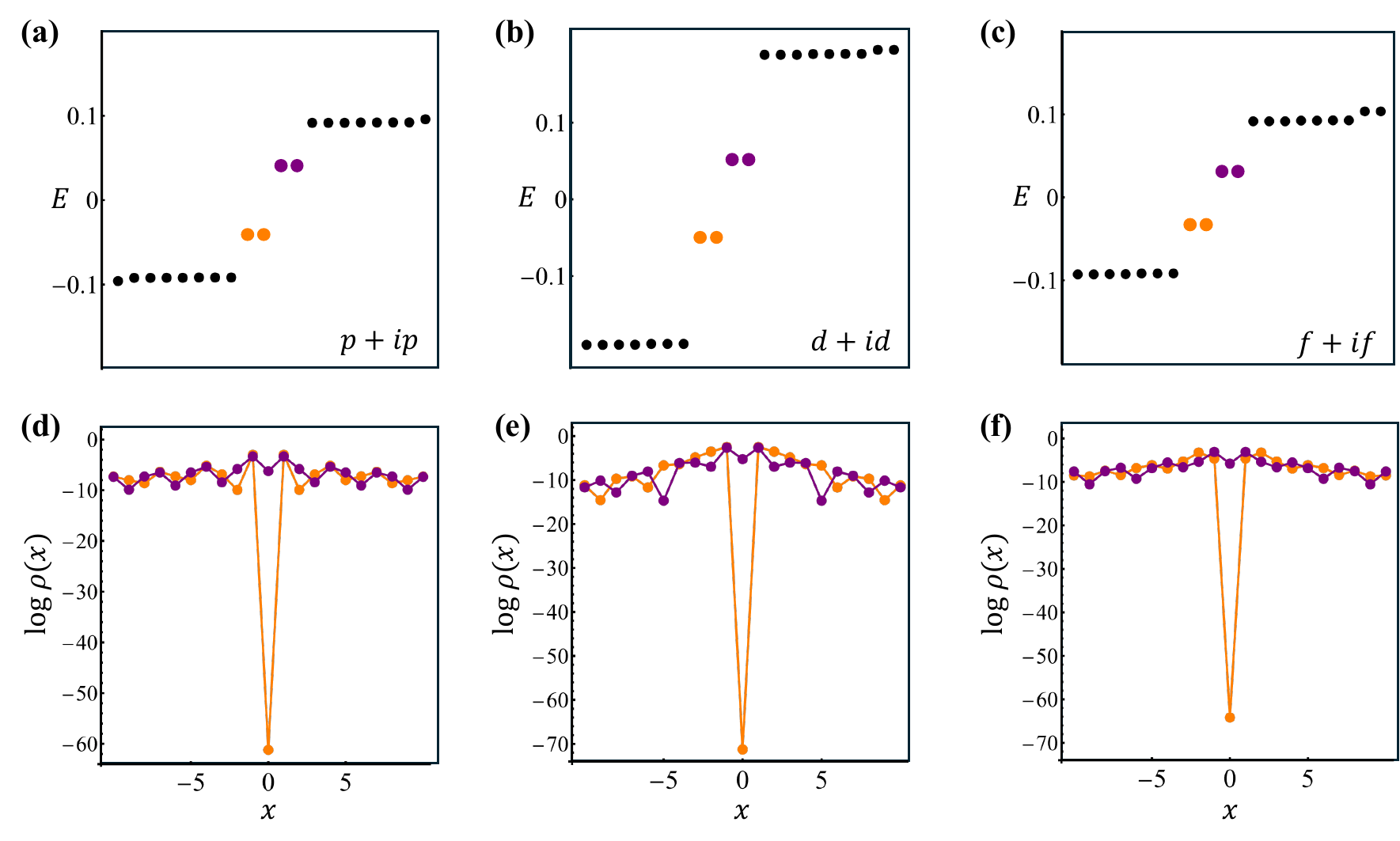}
    \caption{(a–c) Energy spectra for chiral $p+ip$, $d+id$, and $f+if$ superconductors in the presence of a point-like nonmagnetic impurity. Orange and purple dots mark the in-gap impurity-bound states. (d–f) Spatial LDOS maps at the corresponding bound-state energies for chiral $p$-, $d$-, and $f$-wave cases, using the same color scheme as in (a–c). The impurity is located at $x = 0$. All three pairing channels exhibit the predicted node–antinode structure, in agreement with Eq.~\ref{eq:node-antinode relation}.}
    \label{fig:chiral-pdf}
\end{figure*}

We now turn to the bound-state physics induced by a point-like impurity located at ${\bf r} = 0$, modeled by a potential $V \delta({\bf r})$, where ${\bf r} = (r, \theta)$ are in-plane polar coordinates. The impurity can be non-magnetic ($V = U \tau_z$), purely magnetic ($V = M \tau_z \sigma_z$), or a mixture of both. Here, $\tau_{x,y,z}$ are Pauli matrices acting in particle-hole space. As we will show, the detailed type of the impurity is not essential to the main result, provided it induces in-gap bound states. The bound-state energies are determined by solving the matrix equation~\cite{pientka2013topo}
\begin{equation}
    \left[{\bf 1} - V \, {\cal G}(E, {\bf r} = 0)\right] \psi({\bf r} = 0) = 0,
    \label{eq:impurity state equation}
\end{equation}
where $\psi({\bf r})$ is the bound-state wavefunction. The real-space Green’s function at the impurity site is given by ${\cal G}(E, 0) = \int \frac{d^2{\bf k}}{(2\pi)^2} [E - H({\bf k}) + i0^+]^{-1}$. When ${\cal G}(E, {\bf r})$ is analytically known, the spatial profile of the impurity-induced bound state at energy $E_0$ follows as $\psi({\bf r}) = {\cal G}(E_0, {\bf r}) V \psi(0)$.

To proceed analytically, we decompose ${\cal G}(E, 0)$ into a block-diagonal part (normal propagator) and an off-diagonal anomalous part, ${\cal F}(E, 0) \equiv \left[{\cal G}(E, 0)\right]_{12}$, which encodes local pairing information. In the Supplemental Material (SM)~\cite{supp}, we show that each matrix element of the anomalous propagator takes the form ${\cal F}_{ss'}(E, 0) = \frac{1}{(2\pi)^2}\int d^2k F_{ss'}(k,\theta)$, where $s, s' \in \{\uparrow, \downarrow\}$. While the detailed structure of $F_{ss'}(k, \theta)$ depends on the specific electron dispersion and pairing channel, however, we find that $C_n$ symmetry imposes a crucial constraint: $F_{ss'}(k, \theta + 2\pi/n) = F_{ss'}(k, \theta)\text{exp}(i2\pi l/n)$. This $n$-fold periodicity allows us to partition the angular integral into $n$ equal segments, leading to:
\begin{align}
    {\cal F}_{ss'}(E,0)=\begin{cases}
n\int \frac{kdk}{(2\pi)^2}\int_0^{\frac{2\pi}{n}} F_{ss'}(k, \theta)d\theta, & \text{if } l/n\in\mathbb{Z}, \\
0,   & \text{otherwise}.
\end{cases}
\label{eq:anomalous propagator}
\end{align}
Therefore, we have shown that in any $C_n$-invariant single-band superconductor, the anomalous propagator ${\cal F}(E, 0)$ vanishes whenever the pairing phase winding $l$ is an integer multiple of $n$. We note that a special case of Eq.~\ref{eq:anomalous propagator} was previously observed for an isotropic 2d electron-gas model with chiral $p$- and $d$-wave pairings~\cite{wang2004impurity,kaladzhyan2016asymptotic}, although the essential role of crystalline rotation symmetry was not addressed. When ${\cal F}(E, 0) = 0$, the local Green’s function ${\cal G}(E, 0)$ becomes block-diagonal, indicating a complete decoupling between electrons and holes at the impurity site. Consequently, any solution $\psi(0)$ of Eq.~\ref{eq:impurity state equation} must be either purely electron-like or purely hole-like, but not a coherent superposition of both.

Due to particle–hole symmetry, impurity-bound states appear in conjugate pairs, which we formally express as
\begin{equation}
\psi_{\beta+}(0) = (\chi_\beta, 0)^T, \quad \psi_{\beta-}(0) = (0, \chi_\beta^*)^T,
\end{equation}
where $\psi_{\beta+}(0)$ is an electron-like state at energy $E_\beta$, and $\psi_{\beta-}(0)$ is its hole-like partner at $-E_\beta$. Here, $\beta$ labels the impurity state, and $E_\beta$ can be either positive or negative. The local density of states (LDOS) at the impurity site is then given by $\rho_0(E) = \sum_\beta \delta(E - E_\beta) |\chi_\beta|^2$, implying that the LDOS vanishes at one energy (e.g., $-E_\beta$) while remaining finite at the other (e.g., $E_\beta$). We refer to this phenomenon as a ``node–antinode'' structure of the LDOS at the defect center. In a $C_n$-invariant system, this distinctive pattern {\it necessarily} emerges when the following {\it nodal condition} is satisfied:
\begin{equation}
l \not\equiv 0 \pmod{n},
\label{eq:node-antinode relation}
\end{equation}
which follows directly from Eq.~\ref{eq:anomalous propagator}.

In the SM~\cite{supp}, we prove that the nodal condition remains valid for spin-singlet chiral superconductors even in the presence of Rashba spin–orbit coupling. This result is particularly applicable to the Sn/Si(111) system~\cite{ming2023evidence,wu2025Sn}, where chiral $d$-wave pairing and Rashba SOC are both expected to play a role~\cite{rachel2025electronic}. In contrast, Rashba coupling tends to compete with spin-triplet pairing and generally drives the system toward gapless superconductivity. Such scenarios lie beyond the scope of this work, as we focuses on fully gapped chiral states.

{\it Numerical Proof of Nodal Condition} - As proof of concept, we consider a 2d minimal model of chiral superconductor on a square lattice, denoted as $H_{c}({\bf k})$, with electron dispersion $E_{\bf k} = t_x \cos k_x + t_y \cos k_y$ identical for both spin sectors. When $t_x = t_y$, the system respects $C_4$ symmetry. We examine three representative chiral pairing channels: (i) a chiral $p$-wave spin-triplet state with $d_x({\bf k}) = \Delta_0  (\sin k_x+i \sin k_y)$, corresponding to a phase winding number $l = 1$; (ii) a chiral $d$-wave spin-singlet state with $d_y({\bf k}) = 2i \Delta_0 (\cos{k_x}-\cos{k_y}-i\sin{k_x}\sin{k_y})$ and $l = 2$; and (iii) a chiral $f$-wave spin-triplet state with $d_x({\bf k}) = -\Delta_0 [2\sin{k_x}+\cos{k_x}\sin{k_x}-3\cos{k_y}\sin{k_x}-i(2\sin{k_y}-3\cos{k_x}\sin{k_y}+\cos{k_y}\sin{k_y})]$ and $l = 3$. All three pairings satisfy the nodal condition, and we therefore expect each to exhibit the node–antinode LDOS structure in response to a point-like impurity.

We place the BdG Hamiltonian $H({\bf k})$ on a $51 \times 51$ square lattice with parameters $t_x = t_y = 1$, $\mu = 0.5$, and pairing amplitude $\Delta_0 = 0.1$. A nonmagnetic point-like impurity is introduced at the lattice center ${\bf r}_0$ via a local potential $U\tau_z \delta({\bf r}_0)$, where we choose $U_0=10$. To eliminate edge effects and isolate impurity-induced features, we impose periodic boundary conditions in both directions. The resulting impurity spectra for the chiral $p$-, $d$-, and $f$-wave states are shown in Figs. \ref{fig:chiral-pdf} (a)–(c), respectively. In all three cases, we find four in-gap bound states: a spin-degenerate pair at energy $E_0$ and their particle–hole conjugates at $-E_0$. We label the corresponding wavefunctions as $\psi_{\pm, \uparrow/\downarrow}({\bf r})$, where the $\pm$ indicates positive or negative energy. The LDOS is given by $\rho({\bf r}) = \sum_{s=\uparrow,\downarrow} \delta(E - E_0) |\psi_{+,s}({\bf r})|^2 + \delta(E + E_0) |\psi_{-,s}({\bf r})|^2$. 

For the chiral $p$-wave state, Fig. \ref{fig:chiral-pdf} (d) shows a line cut of $\log \rho(x)$ along the $\hat{x}$-axis passing through the impurity site at $x=0$. The LDOS exhibits a pronounced node–antinode structure: at energy $-E_0 = -0.042$ (orange), the LDOS vanishes at the impurity site up to numerical precision ($<10^{-28}$), while it remains finite at $+E_0 = 0.042$ (purple). The same behavior is consistently observed for the chiral $d$- and $f$-wave pairings, as shown in Fig. \ref{fig:chiral-pdf} (e) and (f), respectively. Importantly, this LDOS asymmetry is further resilient against variations of the model parameters. As shown in the SM~\cite{supp}, tuning the chemical potential $\mu$, which modifies both the shape and topological connectivity of the Fermi surfaces, does not alter the node-antinode structure.

\begin{figure}[t]
    \includegraphics[width=0.47\textwidth]{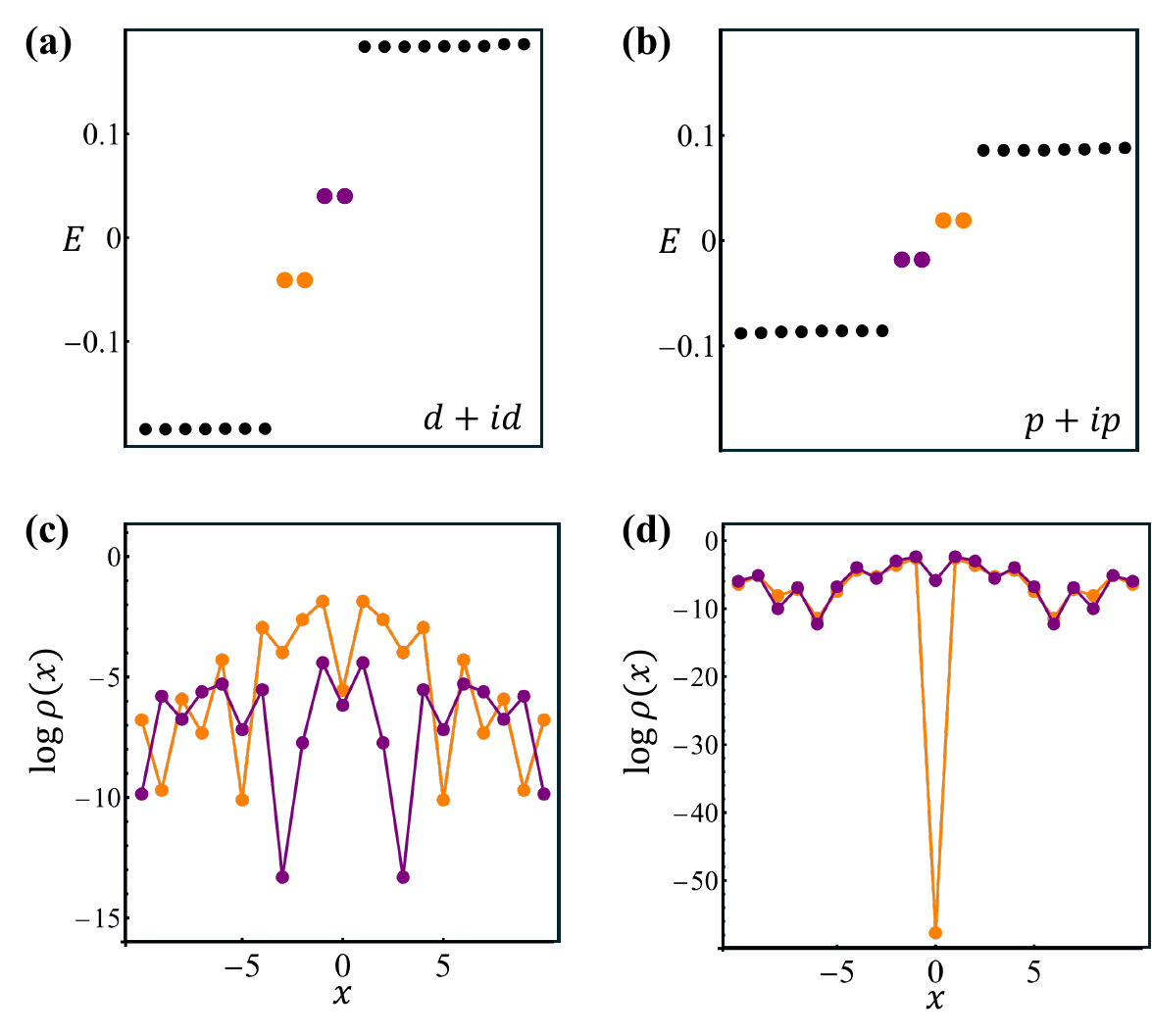}
    \caption{(a–b) Impurity-state spectra with a point-like impurity for chiral $d+id$ and $p+ip$ pairings, respectively, where we reduce the rotation symmetry to $C_2$. For $d+id$ pairing in (c), LDOS plots at both bound-state energies exhibit finite weight at the impurity site, consistent with $l=2\equiv 0$ (mod 2). In contrast, the $p+ip$ case in (d) displays the node-antinode structure since $l=1$.}
    \label{fig:C2}
\end{figure}

A key implication of Eq.~\ref{eq:node-antinode relation} is that the node–antinode LDOS structure disappears when the pairing winding number $l$ is an integer multiple of the crystalline rotation order $n$. To test this exceptional case, we explicitly break $C_4$ symmetry down to $C_2$ by setting $t_x \neq t_y$ in the lattice model $H_c({\bf k})$. In this reduced-symmetry setting, a chiral $d+id$ pairing satisfies $l = n =2$, and hence falls outside the regime where nodal structure is expected. As shown in Fig. \ref{fig:C2} (c), the resulting LDOS no longer exhibits nodal behavior at either positive or negative bound-state energies, in perfect agreement with our prediction. For comparison, we consider chiral $p+ip$ pairing with $l = 1$ in the same $C_2$-symmetric background, where the LDOS retains a clear node-antinode structure, as shown in Fig.~\ref{fig:C2} (d). 

{\it Effects of Extended Impurity} - The derivation of the nodal condition relies on two key simplifying assumptions. First, the impurity is modeled as a point-like $\delta$-function potential, which makes Eq.~\ref{eq:impurity state equation} analytically tractable. In contrast, spatially extended impurities generally complicate the bound-state equations, rendering $\psi(0)$ inaccessible in closed form. Second, our analysis has been restricted to single-band superconductors, where all Fermi surfaces originate from the same Kramers-degenerate band. The situation becomes more intricate in multiband systems, particularly when interband pairing is present. While a general analytic extension of Eq.~\ref{eq:node-antinode relation} to these broader cases appears nontrivial, we test the robustness of the nodal condition against both generalizations through direct numerical simulations below.

\begin{figure}[t]
    \includegraphics[width=0.47\textwidth]{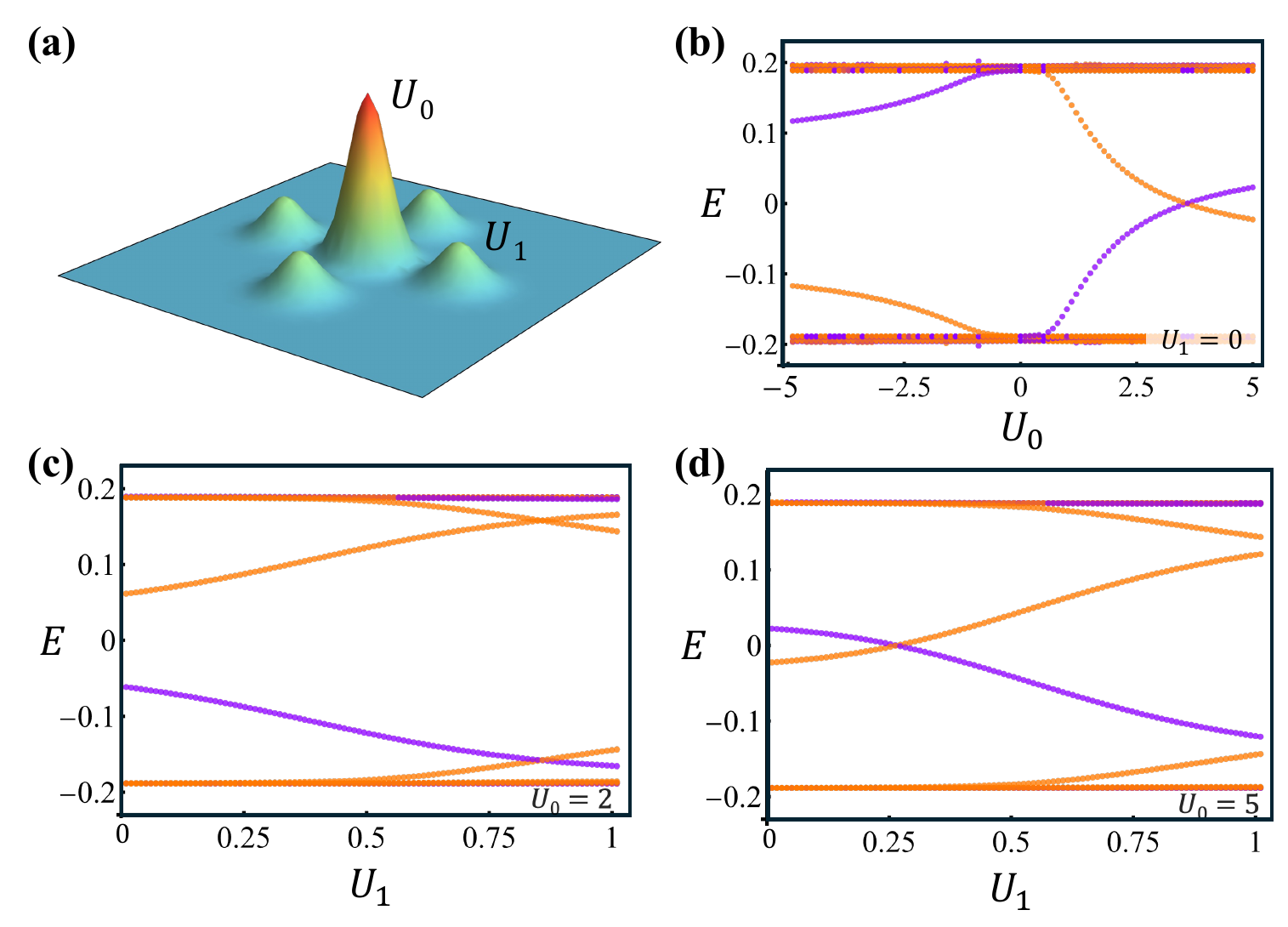}
    \caption{(a) Schematic of an extended impurity potential on a square lattice: the central site hosts an on-site potential $U_0$, surrounded by nearest-neighbor sites with potential $U_1$. (b) Energy spectrum for chiral $d+id$ pairing as a function of $U_0$ with $U_1 = 0$. In-gap impurity states are color-coded by their LDOS contribution at the impurity site: orange denotes vanishing LDOS, while purple indicates finite weight. (c–d) Spectral evolution with increasing $U_1$ at fixed $U_0 = 2$ (c) and $U_0 = 5$ (d). As $U_1$ exceeds the critical value, additional in-gap bound states emerge from the bulk continuum. Both pairs of these new states exhibit vanishing LDOS at the impurity site, as indicated by their orange color.}
    \label{fig:extend-imp}
\end{figure}

We first examine the response of a $C_4$-invariant chiral $d$-wave superconductor, described by $H({\bf k})$, to a spatially extended impurity. As illustrated in Fig. \ref{fig:extend-imp} (a), the impurity potential is modeled by
\begin{equation}
    U({\bf r}) = U_0 \delta({\bf r}_0) + \sum_{\delta{\bf r}} U_1 \delta({\bf r}_0 + \delta{\bf r}),
\end{equation}
where $\delta {\bf r}$ runs over the nearest-neighbor sites surrounding the impurity center ${\bf r}_0$. In the point-like limit ($U_1 = 0$), Fig. \ref{fig:extend-imp} (b) shows the evolution of impurity-bound states as a function of $U_0$. Orange dots indicate ``nodal" BdG eigenstates with vanishing electron amplitude at the impurity site, while purple dots represent finite-amplitude states. Notably, the polarity of the bound states, i.e., whether the positive-energy state is nodal or not, reverses either when the sign of $U_0$ is flipped or when $U_0$ passes through the zero-energy level crossing at $U_0^{(c)} = 2.5$. Despite this phenomenon, the node–antinode structure persists throughout, in agreement with our prediction.

Turning on $U_1$ introduces additional scattering centers that can modify the in-gap spectrum. In Figs.~\ref{fig:extend-imp} (c) and (d), we fix $U_0 = 2$ and $U_0 = 5$, respectively, and plot the bound-state spectra as $U_1$ is gradually increased. The effect of $U_1$ is notably twofold. First, it shifts the energies of the original bound states induced by $U_0$, effectively mimicking a reduction in $U_0$. For instance, in Fig.~\ref{fig:extend-imp} (c), the bound states shift to higher energies with increasing $U_1$, similar to the trend observed when lowering $U_0$ in Fig.~\ref{fig:extend-imp} (b). A comparable behavior appears in Fig.~\ref{fig:extend-imp} (d), where spectral evolution with $U_1$ produces a level crossing akin to that at smaller $U_0$. Crucially, these $U_0$-induced states retain their node–antinode structure across the full range of $U_1$.

Beyond a critical threshold $U_1^{(c)} \approx 0.5t$, two additional pairs of bound states emerge. In contrast to the original states, these $U_1$-induced states exhibit vanishing electron amplitude at the impurity site at both positive and negative energies, resulting in LDOS nodes on both sides of the Fermi level. This ``node–node'' behavior indicates a breakdown of the $\delta$-function approximation, but only for sufficiently extended potentials. In realistic materials where Coulomb potentials are strongly screened, we expect $U_1 \ll U_0$, and the node–antinode structure should remain applicable.

\begin{figure}[t]
    \includegraphics[width=0.47\textwidth]{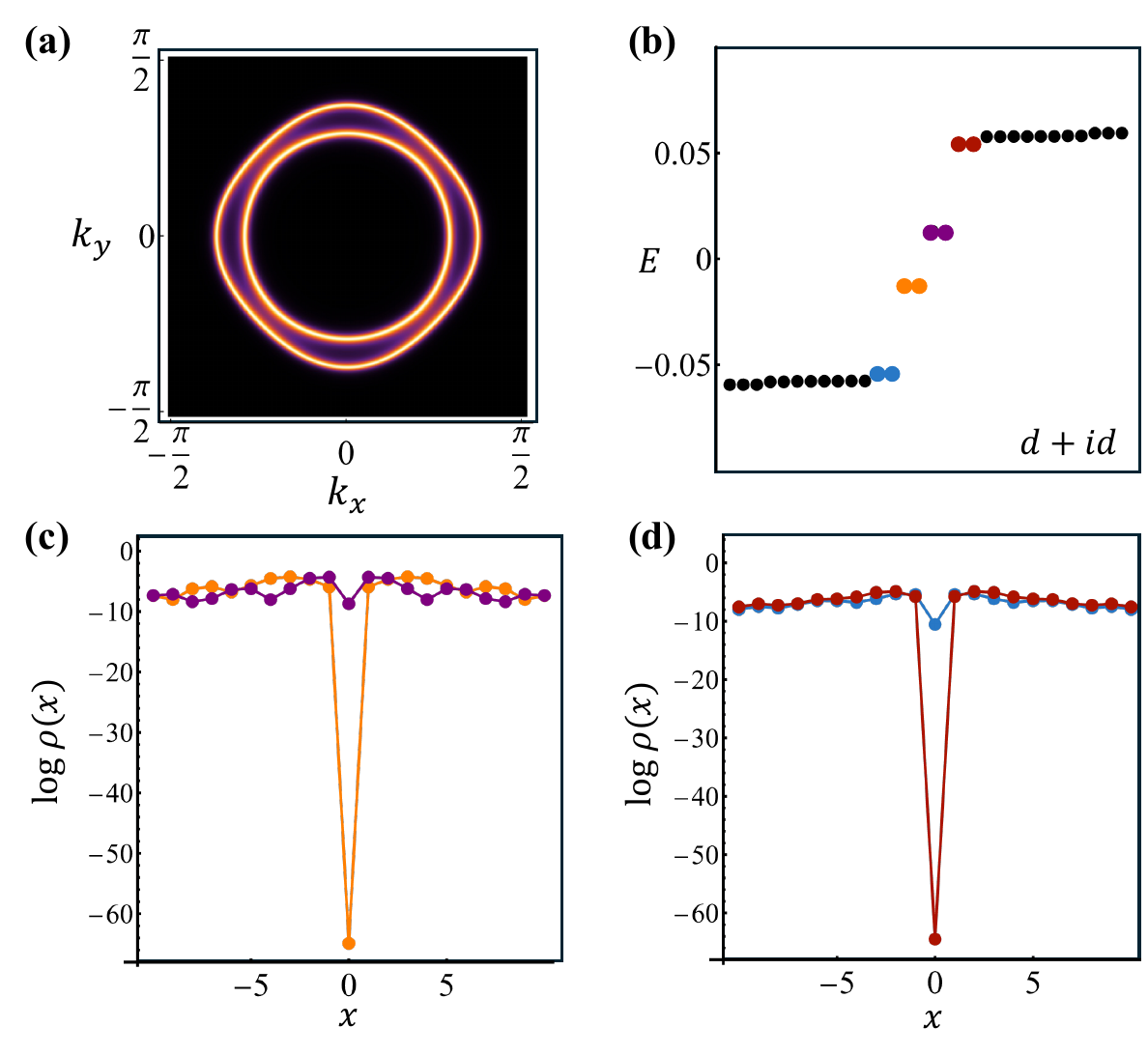}
    \caption{(a) Fermi surface map of the BHZ model at chemical potential $\mu = 0.4$, showing two coexisting Fermi pockets. These give rise to four pairs of in-gap impurity bound states in the presence of chiral $d+id$ pairing, as shown in (b). (c) and (d) display the LDOS profiles for bound states at energies $E = \pm 0.013$ and $E = \pm 0.055$, respectively. Despite the complexity introduced by multiple Fermi surfaces and interband pairing, the characteristic node–antinode structure persists for all bound states.}
    \label{fig:bhz}
\end{figure}

{\it Impurity States in Multiband Systems} - We now explore the validity of the nodal condition in multiband superconductors with interband pairing. To this end, we generalize $H({\bf k})$ to describe a chiral $d$-wave superconductor built from both $s$ and $p_z$ orbitals in the normal state. The electron basis is $|\Psi_e\rangle = (|s,\downarrow\rangle, |s,\uparrow\rangle, |p_z,\uparrow\rangle, |p_z,\downarrow\rangle)^T$. Up to ${\cal O}(k^2)$, the normal-state Hamiltonian takes the Bernevig-Hughes-Zhang (BHZ) form~\cite{bernevig2006quantum},
\begin{equation}
    h_b({\bf k}) = v(\sin k_x\, \kappa_x\sigma_z - \sin k_y\, \kappa_y\sigma_0) + m({\bf k})\, \kappa_z\sigma_0,
\end{equation}
where $\kappa$ and $\sigma$ are Pauli matrices in orbital and spin space, respectively. The mass term is given by $m({\bf k}) = m_0 - m_1(\cos k_x + \cos k_y)$, with $(m_0, m_1) = (3, 2)$ chosen to produce a $\mathbb{Z}_2$ band inversion at the $\Gamma$ point. While the topological character of the bands plays no essential role in the impurity physics discussed below, the band inversion ensures the presence of two spin-degenerate Fermi surfaces with distinct orbital characters. For $\mu = v = 0.4$, both Fermi surfaces are present, as shown in Fig.~\ref{fig:bhz} (a), with the inner and outer contours primarily derived from $p_z$ and $s$ orbitals, respectively.

Building on $h_b({\bf k})$, we introduce a chiral $d$-wave spin-singlet pairing with orbital-dependent pairing strengths. The pairing matrix takes the form $\Delta({\bf k}) = i f({\bf k}) \left( \Delta_s \sigma_y \oplus \Delta_p \sigma_y \right)$, where $f({\bf k})$ is the $d+id$ form factor defined earlier. The spin-orbit coupling parameter $v$ governs the inter-orbital hybridization. As a result, when $\Delta({\bf k})$ is projected onto the band basis, the effective pairing contains both intra- and inter-Fermi-surface components, with the latter scaling with $v$.

Fig.~\ref{fig:bhz} (b) shows the energy spectrum of the multiband system on a $51\times51$ lattice in the presence of a point-like impurity with strength $U_0 = 40$. We set $(\Delta_s, \Delta_p) = (0.26, 0.14)$, while all other parameters match those used to generate the Fermi surfaces in Fig.~\ref{fig:bhz} (a). Compared to the single-band case in Fig.~\ref{fig:chiral-pdf}, the number of Fermi surfaces doubles, and accordingly we observe four pairs of in-gap impurity-bound states, each Fermi surface contributing one particle–hole-conjugate pair. The spatial LDOS profiles of the bound states $E = \pm 0.013$ (in purple and orange) and $E = \pm 0.055$ (in red and blue) are shown in Figs.~\ref{fig:bhz} (c) and (d), respectively. Remarkably, the LDOS exhibits a clear node–antinode asymmetry at opposite bound-state energies. These findings offer strong numerical support that the nodal condition, originally derived for single-band systems, continues to hold in multiband chiral superconductors.

{\it Discussions} - We have established a new real-space diagnostic for chiral superconductivity: a robust node–antinode structure in the LDOS induced by point-like impurities. This striking pattern arises from a general symmetry principle, the nodal condition, which connects the phase winding of chiral Cooper pairs with the crystalline point-group symmetry of the host lattice. Unlike conventional probes that focus on global signatures of time-reversal symmetry breaking, our framework identifies a direct, microscopic fingerprint of chiral pairing that is directly accessible with STM technique. 

This framework has already demonstrated its power in understanding recent high-resolution STM measurements on Sn/Si(111)~\cite{wu2025Sn}, a 2d single-band triangular-lattice system believed to host chiral $d$-wave superconductivity. The observed LDOS patterns around atomic-scale defects unambiguously exhibit a node–antinode structure, in precise agreement with the nodal condition. This provides strong support for the chiral $d$-wave scenario. The success of this analysis underscores the practical utility of our theory in realistic materials and further motivates its future application to other candidate systems~\cite{ran2019nearly,jiao2020chiral,schemm2014observation,avers2020broken,ribak2020chiral,zhao2023time,wan2024unconventional,han2025signatures}, where the chiral nature of pairing symmetry remains an open question.

It is important to emphasize that while the nodal condition is a necessary criterion for the emergence of the node–antinode LDOS structure, it is {\it not sufficient}. In principle, fine-tuned nonchiral superconductors or impurity configurations may also yield vanishing LDOS at specific energies. However, such scenarios are fragile and lack robustness to perturbations. In contrast, similar LDOS pattern in chiral superconductors should be inherent and stable across variations in system parameters. Therefore, one can, in practice, examine whether the LDOS nodal structure persists across multiple types of defects in a given material, which will provide further evidence to prove or defy the possibility of chiral Cooper pairing. For example, in Sn/Si(111), the same node–antinode pattern has been consistently observed around various impurity configurations~\cite{wu2025Sn}, strongly supporting its intrinsic and chiral origin.

Finally, it would be interesting to extend our framework to other real-space defects, particularly topological ones such as lattice dislocations~\cite{asahi2012defect,hughes2014dislocation,hu2024dislocation,breio2024dislocation} and vortices~\cite{fu2008vortex,hosur2011vortex,hu2022vortex,hu2023topological,holmvall2023coreless,holmvall2023robust}. We expect that the interplay among lattice symmetry, the real-space topology of defects~\cite{mermin1979RMP}, and the $k$-space winding of Cooper pairs will reveal new and intriguing signatures of chiral superconductivity. Moreover, our work opens a fresh perspective on how microscopic pair-breaking processes decode the nature of Cooper pairs. For example, extending this framework to non-chiral superconductors with pairing symmetries that transform nontrivially under other lattice operations, such as reflections, may uncover new, distinctive defect responses. We leave these promising directions for future exploration.

{\it Acknowledgement -} We thank Z. Chen, S. Johnston, F. Ming, and especially H.H. Weitering for insightful discussions. This work is supported by a start-up fund at the University of Tennessee.

\bibliography{ref}

\begin{thebibliography}{42}%
\makeatletter
\providecommand \@ifxundefined [1]{%
 \@ifx{#1\undefined}
}%
\providecommand \@ifnum [1]{%
 \ifnum #1\expandafter \@firstoftwo
 \else \expandafter \@secondoftwo
 \fi
}%
\providecommand \@ifx [1]{%
 \ifx #1\expandafter \@firstoftwo
 \else \expandafter \@secondoftwo
 \fi
}%
\providecommand \natexlab [1]{#1}%
\providecommand \enquote  [1]{``#1''}%
\providecommand \bibnamefont  [1]{#1}%
\providecommand \bibfnamefont [1]{#1}%
\providecommand \citenamefont [1]{#1}%
\providecommand \href@noop [0]{\@secondoftwo}%
\providecommand \href [0]{\begingroup \@sanitize@url \@href}%
\providecommand \@href[1]{\@@startlink{#1}\@@href}%
\providecommand \@@href[1]{\endgroup#1\@@endlink}%
\providecommand \@sanitize@url [0]{\catcode `\\12\catcode `\$12\catcode
  `\&12\catcode `\#12\catcode `\^12\catcode `\_12\catcode `\%12\relax}%
\providecommand \@@startlink[1]{}%
\providecommand \@@endlink[0]{}%
\providecommand \url  [0]{\begingroup\@sanitize@url \@url }%
\providecommand \@url [1]{\endgroup\@href {#1}{\urlprefix }}%
\providecommand \urlprefix  [0]{URL }%
\providecommand \Eprint [0]{\href }%
\providecommand \doibase [0]{https://doi.org/}%
\providecommand \selectlanguage [0]{\@gobble}%
\providecommand \bibinfo  [0]{\@secondoftwo}%
\providecommand \bibfield  [0]{\@secondoftwo}%
\providecommand \translation [1]{[#1]}%
\providecommand \BibitemOpen [0]{}%
\providecommand \bibitemStop [0]{}%
\providecommand \bibitemNoStop [0]{.\EOS\space}%
\providecommand \EOS [0]{\spacefactor3000\relax}%
\providecommand \BibitemShut  [1]{\csname bibitem#1\endcsname}%
\let\auto@bib@innerbib\@empty
\bibitem [{\citenamefont {Read}\ and\ \citenamefont
  {Green}(2000)}]{read2000paired}%
  \BibitemOpen
  \bibfield  {author} {\bibinfo {author} {\bibfnamefont {N.}~\bibnamefont
  {Read}}\ and\ \bibinfo {author} {\bibfnamefont {D.}~\bibnamefont {Green}},\
  }\bibfield  {title} {\bibinfo {title} {Paired states of fermions in two
  dimensions with breaking of parity and time-reversal symmetries and the
  fractional quantum hall effect},\ }\href
  {https://doi.org/10.1103/PhysRevB.61.10267} {\bibfield  {journal} {\bibinfo
  {journal} {Phys. Rev. B}\ }\textbf {\bibinfo {volume} {61}},\ \bibinfo
  {pages} {10267} (\bibinfo {year} {2000})}\BibitemShut {NoStop}%
\bibitem [{\citenamefont {Ivanov}(2001)}]{ivanov2001nonabelian}%
  \BibitemOpen
  \bibfield  {author} {\bibinfo {author} {\bibfnamefont {D.~A.}\ \bibnamefont
  {Ivanov}},\ }\bibfield  {title} {\bibinfo {title} {Non-abelian statistics of
  half-quantum vortices in $\mathit{p}$-wave superconductors},\ }\href
  {https://doi.org/10.1103/PhysRevLett.86.268} {\bibfield  {journal} {\bibinfo
  {journal} {Phys. Rev. Lett.}\ }\textbf {\bibinfo {volume} {86}},\ \bibinfo
  {pages} {268} (\bibinfo {year} {2001})}\BibitemShut {NoStop}%
\bibitem [{\citenamefont {Sau}\ \emph {et~al.}(2010)\citenamefont {Sau},
  \citenamefont {Lutchyn}, \citenamefont {Tewari},\ and\ \citenamefont
  {Das~Sarma}}]{sau2010generic}%
  \BibitemOpen
  \bibfield  {author} {\bibinfo {author} {\bibfnamefont {J.~D.}\ \bibnamefont
  {Sau}}, \bibinfo {author} {\bibfnamefont {R.~M.}\ \bibnamefont {Lutchyn}},
  \bibinfo {author} {\bibfnamefont {S.}~\bibnamefont {Tewari}},\ and\ \bibinfo
  {author} {\bibfnamefont {S.}~\bibnamefont {Das~Sarma}},\ }\bibfield  {title}
  {\bibinfo {title} {Generic new platform for topological quantum computation
  using semiconductor heterostructures},\ }\href
  {https://doi.org/10.1103/PhysRevLett.104.040502} {\bibfield  {journal}
  {\bibinfo  {journal} {Phys. Rev. Lett.}\ }\textbf {\bibinfo {volume} {104}},\
  \bibinfo {pages} {040502} (\bibinfo {year} {2010})}\BibitemShut {NoStop}%
\bibitem [{\citenamefont {Qi}\ \emph {et~al.}(2010)\citenamefont {Qi},
  \citenamefont {Hughes},\ and\ \citenamefont {Zhang}}]{qi2010chiral}%
  \BibitemOpen
  \bibfield  {author} {\bibinfo {author} {\bibfnamefont {X.-L.}\ \bibnamefont
  {Qi}}, \bibinfo {author} {\bibfnamefont {T.~L.}\ \bibnamefont {Hughes}},\
  and\ \bibinfo {author} {\bibfnamefont {S.-C.}\ \bibnamefont {Zhang}},\
  }\bibfield  {title} {\bibinfo {title} {Chiral topological superconductor from
  the quantum hall state},\ }\href {https://doi.org/10.1103/PhysRevB.82.184516}
  {\bibfield  {journal} {\bibinfo  {journal} {Phys. Rev. B}\ }\textbf {\bibinfo
  {volume} {82}},\ \bibinfo {pages} {184516} (\bibinfo {year}
  {2010})}\BibitemShut {NoStop}%
\bibitem [{\citenamefont {Kallin}\ and\ \citenamefont
  {Berlinsky}(2016)}]{kallin2016chiral}%
  \BibitemOpen
  \bibfield  {author} {\bibinfo {author} {\bibfnamefont {C.}~\bibnamefont
  {Kallin}}\ and\ \bibinfo {author} {\bibfnamefont {J.}~\bibnamefont
  {Berlinsky}},\ }\bibfield  {title} {\bibinfo {title} {Chiral
  superconductors},\ }\href
  {https://iopscience.iop.org/article/10.1088/0034-4885/79/5/054502} {\bibfield
   {journal} {\bibinfo  {journal} {Reports on Progress in Physics}\ }\textbf
  {\bibinfo {volume} {79}},\ \bibinfo {pages} {054502} (\bibinfo {year}
  {2016})}\BibitemShut {NoStop}%
\bibitem [{\citenamefont {Can}\ \emph {et~al.}(2021)\citenamefont {Can},
  \citenamefont {Tummuru}, \citenamefont {Day}, \citenamefont {Elfimov},
  \citenamefont {Damascelli},\ and\ \citenamefont {Franz}}]{can2021high}%
  \BibitemOpen
  \bibfield  {author} {\bibinfo {author} {\bibfnamefont {O.}~\bibnamefont
  {Can}}, \bibinfo {author} {\bibfnamefont {T.}~\bibnamefont {Tummuru}},
  \bibinfo {author} {\bibfnamefont {R.~P.}\ \bibnamefont {Day}}, \bibinfo
  {author} {\bibfnamefont {I.}~\bibnamefont {Elfimov}}, \bibinfo {author}
  {\bibfnamefont {A.}~\bibnamefont {Damascelli}},\ and\ \bibinfo {author}
  {\bibfnamefont {M.}~\bibnamefont {Franz}},\ }\bibfield  {title} {\bibinfo
  {title} {High-temperature topological superconductivity in twisted
  double-layer copper oxides},\ }\href
  {https://doi.org/10.1038/s41567-020-01142-7} {\bibfield  {journal} {\bibinfo
  {journal} {Nature Physics}\ }\textbf {\bibinfo {volume} {17}},\ \bibinfo
  {pages} {519} (\bibinfo {year} {2021})}\BibitemShut {NoStop}%
\bibitem [{\citenamefont {Volovik}(1999)}]{volovik1999fermion}%
  \BibitemOpen
  \bibfield  {author} {\bibinfo {author} {\bibfnamefont {G.}~\bibnamefont
  {Volovik}},\ }\bibfield  {title} {\bibinfo {title} {Fermion zero modes on
  vortices in chiral superconductors},\ }\href
  {https://doi.org/10.1134/1.568223} {\bibfield  {journal} {\bibinfo  {journal}
  {Journal of Experimental and Theoretical Physics Letters}\ }\textbf {\bibinfo
  {volume} {70}},\ \bibinfo {pages} {609} (\bibinfo {year} {1999})}\BibitemShut
  {NoStop}%
\bibitem [{\citenamefont {Nayak}\ \emph {et~al.}(2008)\citenamefont {Nayak},
  \citenamefont {Simon}, \citenamefont {Stern}, \citenamefont {Freedman},\ and\
  \citenamefont {Das~Sarma}}]{nayak2008RMP}%
  \BibitemOpen
  \bibfield  {author} {\bibinfo {author} {\bibfnamefont {C.}~\bibnamefont
  {Nayak}}, \bibinfo {author} {\bibfnamefont {S.~H.}\ \bibnamefont {Simon}},
  \bibinfo {author} {\bibfnamefont {A.}~\bibnamefont {Stern}}, \bibinfo
  {author} {\bibfnamefont {M.}~\bibnamefont {Freedman}},\ and\ \bibinfo
  {author} {\bibfnamefont {S.}~\bibnamefont {Das~Sarma}},\ }\bibfield  {title}
  {\bibinfo {title} {Non-abelian anyons and topological quantum computation},\
  }\href {https://doi.org/10.1103/RevModPhys.80.1083} {\bibfield  {journal}
  {\bibinfo  {journal} {Rev. Mod. Phys.}\ }\textbf {\bibinfo {volume} {80}},\
  \bibinfo {pages} {1083} (\bibinfo {year} {2008})}\BibitemShut {NoStop}%
\bibitem [{\citenamefont {Luke}\ \emph {et~al.}(1998)\citenamefont {Luke},
  \citenamefont {Fudamoto}, \citenamefont {Kojima}, \citenamefont {Larkin},
  \citenamefont {Merrin}, \citenamefont {Nachumi}, \citenamefont {Uemura},
  \citenamefont {Maeno}, \citenamefont {Mao}, \citenamefont {Mori} \emph
  {et~al.}}]{luke1998time}%
  \BibitemOpen
  \bibfield  {author} {\bibinfo {author} {\bibfnamefont {G.~M.}\ \bibnamefont
  {Luke}}, \bibinfo {author} {\bibfnamefont {Y.}~\bibnamefont {Fudamoto}},
  \bibinfo {author} {\bibfnamefont {K.}~\bibnamefont {Kojima}}, \bibinfo
  {author} {\bibfnamefont {M.}~\bibnamefont {Larkin}}, \bibinfo {author}
  {\bibfnamefont {J.}~\bibnamefont {Merrin}}, \bibinfo {author} {\bibfnamefont
  {B.}~\bibnamefont {Nachumi}}, \bibinfo {author} {\bibfnamefont
  {Y.}~\bibnamefont {Uemura}}, \bibinfo {author} {\bibfnamefont
  {Y.}~\bibnamefont {Maeno}}, \bibinfo {author} {\bibfnamefont
  {Z.}~\bibnamefont {Mao}}, \bibinfo {author} {\bibfnamefont {Y.}~\bibnamefont
  {Mori}}, \emph {et~al.},\ }\bibfield  {title} {\bibinfo {title}
  {Time-reversal symmetry-breaking superconductivity in sr2ruo4},\ }\href
  {https://doi.org/10.1038/29038} {\bibfield  {journal} {\bibinfo  {journal}
  {Nature}\ }\textbf {\bibinfo {volume} {394}},\ \bibinfo {pages} {558}
  (\bibinfo {year} {1998})}\BibitemShut {NoStop}%
\bibitem [{\citenamefont {Mackenzie}\ and\ \citenamefont
  {Maeno}(2003)}]{mackenzie2003RMP}%
  \BibitemOpen
  \bibfield  {author} {\bibinfo {author} {\bibfnamefont {A.~P.}\ \bibnamefont
  {Mackenzie}}\ and\ \bibinfo {author} {\bibfnamefont {Y.}~\bibnamefont
  {Maeno}},\ }\bibfield  {title} {\bibinfo {title} {The superconductivity of
  ${\mathrm{sr}}_{2}{\mathrm{ruo}}_{4}$ and the physics of spin-triplet
  pairing},\ }\href {https://doi.org/10.1103/RevModPhys.75.657} {\bibfield
  {journal} {\bibinfo  {journal} {Rev. Mod. Phys.}\ }\textbf {\bibinfo {volume}
  {75}},\ \bibinfo {pages} {657} (\bibinfo {year} {2003})}\BibitemShut
  {NoStop}%
\bibitem [{\citenamefont {Ran}\ \emph {et~al.}(2019)\citenamefont {Ran},
  \citenamefont {Eckberg}, \citenamefont {Ding}, \citenamefont {Furukawa},
  \citenamefont {Metz}, \citenamefont {Saha}, \citenamefont {Liu},
  \citenamefont {Zic}, \citenamefont {Kim}, \citenamefont {Paglione} \emph
  {et~al.}}]{ran2019nearly}%
  \BibitemOpen
  \bibfield  {author} {\bibinfo {author} {\bibfnamefont {S.}~\bibnamefont
  {Ran}}, \bibinfo {author} {\bibfnamefont {C.}~\bibnamefont {Eckberg}},
  \bibinfo {author} {\bibfnamefont {Q.-P.}\ \bibnamefont {Ding}}, \bibinfo
  {author} {\bibfnamefont {Y.}~\bibnamefont {Furukawa}}, \bibinfo {author}
  {\bibfnamefont {T.}~\bibnamefont {Metz}}, \bibinfo {author} {\bibfnamefont
  {S.~R.}\ \bibnamefont {Saha}}, \bibinfo {author} {\bibfnamefont {I.-L.}\
  \bibnamefont {Liu}}, \bibinfo {author} {\bibfnamefont {M.}~\bibnamefont
  {Zic}}, \bibinfo {author} {\bibfnamefont {H.}~\bibnamefont {Kim}}, \bibinfo
  {author} {\bibfnamefont {J.}~\bibnamefont {Paglione}}, \emph {et~al.},\
  }\bibfield  {title} {\bibinfo {title} {Nearly ferromagnetic spin-triplet
  superconductivity},\ }\href
  {https://www.science.org/doi/10.1126/science.aav8645} {\bibfield  {journal}
  {\bibinfo  {journal} {Science}\ }\textbf {\bibinfo {volume} {365}},\ \bibinfo
  {pages} {684} (\bibinfo {year} {2019})}\BibitemShut {NoStop}%
\bibitem [{\citenamefont {Jiao}\ \emph {et~al.}(2020)\citenamefont {Jiao},
  \citenamefont {Howard}, \citenamefont {Ran}, \citenamefont {Wang},
  \citenamefont {Rodriguez}, \citenamefont {Sigrist}, \citenamefont {Wang},
  \citenamefont {Butch},\ and\ \citenamefont {Madhavan}}]{jiao2020chiral}%
  \BibitemOpen
  \bibfield  {author} {\bibinfo {author} {\bibfnamefont {L.}~\bibnamefont
  {Jiao}}, \bibinfo {author} {\bibfnamefont {S.}~\bibnamefont {Howard}},
  \bibinfo {author} {\bibfnamefont {S.}~\bibnamefont {Ran}}, \bibinfo {author}
  {\bibfnamefont {Z.}~\bibnamefont {Wang}}, \bibinfo {author} {\bibfnamefont
  {J.~O.}\ \bibnamefont {Rodriguez}}, \bibinfo {author} {\bibfnamefont
  {M.}~\bibnamefont {Sigrist}}, \bibinfo {author} {\bibfnamefont
  {Z.}~\bibnamefont {Wang}}, \bibinfo {author} {\bibfnamefont {N.~P.}\
  \bibnamefont {Butch}},\ and\ \bibinfo {author} {\bibfnamefont
  {V.}~\bibnamefont {Madhavan}},\ }\bibfield  {title} {\bibinfo {title} {Chiral
  superconductivity in heavy-fermion metal ute2},\ }\href
  {https://doi.org/10.1038/s41586-020-2122-2} {\bibfield  {journal} {\bibinfo
  {journal} {Nature}\ }\textbf {\bibinfo {volume} {579}},\ \bibinfo {pages}
  {523} (\bibinfo {year} {2020})}\BibitemShut {NoStop}%
\bibitem [{\citenamefont {Schemm}\ \emph {et~al.}(2014)\citenamefont {Schemm},
  \citenamefont {Gannon}, \citenamefont {Wishne}, \citenamefont {Halperin},\
  and\ \citenamefont {Kapitulnik}}]{schemm2014observation}%
  \BibitemOpen
  \bibfield  {author} {\bibinfo {author} {\bibfnamefont {E.}~\bibnamefont
  {Schemm}}, \bibinfo {author} {\bibfnamefont {W.}~\bibnamefont {Gannon}},
  \bibinfo {author} {\bibfnamefont {C.}~\bibnamefont {Wishne}}, \bibinfo
  {author} {\bibfnamefont {W.}~\bibnamefont {Halperin}},\ and\ \bibinfo
  {author} {\bibfnamefont {A.}~\bibnamefont {Kapitulnik}},\ }\bibfield  {title}
  {\bibinfo {title} {Observation of broken time-reversal symmetry in the
  heavy-fermion superconductor upt3},\ }\href
  {https://www.science.org/doi/10.1126/science.1248552} {\bibfield  {journal}
  {\bibinfo  {journal} {Science}\ }\textbf {\bibinfo {volume} {345}},\ \bibinfo
  {pages} {190} (\bibinfo {year} {2014})}\BibitemShut {NoStop}%
\bibitem [{\citenamefont {Avers}\ \emph {et~al.}(2020)\citenamefont {Avers},
  \citenamefont {Gannon}, \citenamefont {Kuhn}, \citenamefont {Halperin},
  \citenamefont {Sauls}, \citenamefont {DeBeer-Schmitt}, \citenamefont
  {Dewhurst}, \citenamefont {Gavilano}, \citenamefont {Nagy}, \citenamefont
  {Gasser} \emph {et~al.}}]{avers2020broken}%
  \BibitemOpen
  \bibfield  {author} {\bibinfo {author} {\bibfnamefont {K.~E.}\ \bibnamefont
  {Avers}}, \bibinfo {author} {\bibfnamefont {W.~J.}\ \bibnamefont {Gannon}},
  \bibinfo {author} {\bibfnamefont {S.~J.}\ \bibnamefont {Kuhn}}, \bibinfo
  {author} {\bibfnamefont {W.~P.}\ \bibnamefont {Halperin}}, \bibinfo {author}
  {\bibfnamefont {J.}~\bibnamefont {Sauls}}, \bibinfo {author} {\bibfnamefont
  {L.}~\bibnamefont {DeBeer-Schmitt}}, \bibinfo {author} {\bibfnamefont
  {C.}~\bibnamefont {Dewhurst}}, \bibinfo {author} {\bibfnamefont
  {J.}~\bibnamefont {Gavilano}}, \bibinfo {author} {\bibfnamefont
  {G.}~\bibnamefont {Nagy}}, \bibinfo {author} {\bibfnamefont {U.}~\bibnamefont
  {Gasser}}, \emph {et~al.},\ }\bibfield  {title} {\bibinfo {title} {Broken
  time-reversal symmetry in the topological superconductor upt3},\ }\href
  {https://doi.org/10.1038/s41567-020-0822-z} {\bibfield  {journal} {\bibinfo
  {journal} {Nature Physics}\ }\textbf {\bibinfo {volume} {16}},\ \bibinfo
  {pages} {531} (\bibinfo {year} {2020})}\BibitemShut {NoStop}%
\bibitem [{\citenamefont {Ribak}\ \emph {et~al.}(2020)\citenamefont {Ribak},
  \citenamefont {Skiff}, \citenamefont {Mograbi}, \citenamefont {Rout},
  \citenamefont {Fischer}, \citenamefont {Ruhman}, \citenamefont {Chashka},
  \citenamefont {Dagan},\ and\ \citenamefont {Kanigel}}]{ribak2020chiral}%
  \BibitemOpen
  \bibfield  {author} {\bibinfo {author} {\bibfnamefont {A.}~\bibnamefont
  {Ribak}}, \bibinfo {author} {\bibfnamefont {R.~M.}\ \bibnamefont {Skiff}},
  \bibinfo {author} {\bibfnamefont {M.}~\bibnamefont {Mograbi}}, \bibinfo
  {author} {\bibfnamefont {P.}~\bibnamefont {Rout}}, \bibinfo {author}
  {\bibfnamefont {M.}~\bibnamefont {Fischer}}, \bibinfo {author} {\bibfnamefont
  {J.}~\bibnamefont {Ruhman}}, \bibinfo {author} {\bibfnamefont
  {K.}~\bibnamefont {Chashka}}, \bibinfo {author} {\bibfnamefont
  {Y.}~\bibnamefont {Dagan}},\ and\ \bibinfo {author} {\bibfnamefont
  {A.}~\bibnamefont {Kanigel}},\ }\bibfield  {title} {\bibinfo {title} {Chiral
  superconductivity in the alternate stacking compound 4hb-tas2},\ }\href
  {https://www.science.org/doi/10.1126/sciadv.aax9480} {\bibfield  {journal}
  {\bibinfo  {journal} {Science advances}\ }\textbf {\bibinfo {volume} {6}},\
  \bibinfo {pages} {eaax9480} (\bibinfo {year} {2020})}\BibitemShut {NoStop}%
\bibitem [{\citenamefont {Zhao}\ \emph {et~al.}(2023)\citenamefont {Zhao},
  \citenamefont {Cui}, \citenamefont {Volkov}, \citenamefont {Yoo},
  \citenamefont {Lee}, \citenamefont {Gardener}, \citenamefont {Akey},
  \citenamefont {Engelke}, \citenamefont {Ronen}, \citenamefont {Zhong} \emph
  {et~al.}}]{zhao2023time}%
  \BibitemOpen
  \bibfield  {author} {\bibinfo {author} {\bibfnamefont {S.~F.}\ \bibnamefont
  {Zhao}}, \bibinfo {author} {\bibfnamefont {X.}~\bibnamefont {Cui}}, \bibinfo
  {author} {\bibfnamefont {P.~A.}\ \bibnamefont {Volkov}}, \bibinfo {author}
  {\bibfnamefont {H.}~\bibnamefont {Yoo}}, \bibinfo {author} {\bibfnamefont
  {S.}~\bibnamefont {Lee}}, \bibinfo {author} {\bibfnamefont {J.~A.}\
  \bibnamefont {Gardener}}, \bibinfo {author} {\bibfnamefont {A.~J.}\
  \bibnamefont {Akey}}, \bibinfo {author} {\bibfnamefont {R.}~\bibnamefont
  {Engelke}}, \bibinfo {author} {\bibfnamefont {Y.}~\bibnamefont {Ronen}},
  \bibinfo {author} {\bibfnamefont {R.}~\bibnamefont {Zhong}}, \emph {et~al.},\
  }\bibfield  {title} {\bibinfo {title} {Time-reversal symmetry breaking
  superconductivity between twisted cuprate superconductors},\ }\href
  {https://www.science.org/doi/10.1126/science.abl8371} {\bibfield  {journal}
  {\bibinfo  {journal} {Science}\ }\textbf {\bibinfo {volume} {382}},\ \bibinfo
  {pages} {1422} (\bibinfo {year} {2023})}\BibitemShut {NoStop}%
\bibitem [{\citenamefont {Wan}\ \emph {et~al.}(2024)\citenamefont {Wan},
  \citenamefont {Qiu}, \citenamefont {Ren}, \citenamefont {Qian}, \citenamefont
  {Li}, \citenamefont {Xu}, \citenamefont {Zhou}, \citenamefont {Zhou},
  \citenamefont {Zhou}, \citenamefont {Wang} \emph
  {et~al.}}]{wan2024unconventional}%
  \BibitemOpen
  \bibfield  {author} {\bibinfo {author} {\bibfnamefont {Z.}~\bibnamefont
  {Wan}}, \bibinfo {author} {\bibfnamefont {G.}~\bibnamefont {Qiu}}, \bibinfo
  {author} {\bibfnamefont {H.}~\bibnamefont {Ren}}, \bibinfo {author}
  {\bibfnamefont {Q.}~\bibnamefont {Qian}}, \bibinfo {author} {\bibfnamefont
  {Y.}~\bibnamefont {Li}}, \bibinfo {author} {\bibfnamefont {D.}~\bibnamefont
  {Xu}}, \bibinfo {author} {\bibfnamefont {J.}~\bibnamefont {Zhou}}, \bibinfo
  {author} {\bibfnamefont {J.}~\bibnamefont {Zhou}}, \bibinfo {author}
  {\bibfnamefont {B.}~\bibnamefont {Zhou}}, \bibinfo {author} {\bibfnamefont
  {L.}~\bibnamefont {Wang}}, \emph {et~al.},\ }\bibfield  {title} {\bibinfo
  {title} {Unconventional superconductivity in chiral molecule--tas2 hybrid
  superlattices},\ }\href {https://doi.org/10.1038/s41586-024-07625-4}
  {\bibfield  {journal} {\bibinfo  {journal} {Nature}\ }\textbf {\bibinfo
  {volume} {632}},\ \bibinfo {pages} {69} (\bibinfo {year} {2024})}\BibitemShut
  {NoStop}%
\bibitem [{\citenamefont {Han}\ \emph {et~al.}(2025)\citenamefont {Han},
  \citenamefont {Lu}, \citenamefont {Hadjri}, \citenamefont {Shi},
  \citenamefont {Wu}, \citenamefont {Xu}, \citenamefont {Yao}, \citenamefont
  {Cotten}, \citenamefont {Sedeh}, \citenamefont {Weldeyesus} \emph
  {et~al.}}]{han2025signatures}%
  \BibitemOpen
  \bibfield  {author} {\bibinfo {author} {\bibfnamefont {T.}~\bibnamefont
  {Han}}, \bibinfo {author} {\bibfnamefont {Z.}~\bibnamefont {Lu}}, \bibinfo
  {author} {\bibfnamefont {Z.}~\bibnamefont {Hadjri}}, \bibinfo {author}
  {\bibfnamefont {L.}~\bibnamefont {Shi}}, \bibinfo {author} {\bibfnamefont
  {Z.}~\bibnamefont {Wu}}, \bibinfo {author} {\bibfnamefont {W.}~\bibnamefont
  {Xu}}, \bibinfo {author} {\bibfnamefont {Y.}~\bibnamefont {Yao}}, \bibinfo
  {author} {\bibfnamefont {A.~A.}\ \bibnamefont {Cotten}}, \bibinfo {author}
  {\bibfnamefont {O.~S.}\ \bibnamefont {Sedeh}}, \bibinfo {author}
  {\bibfnamefont {H.}~\bibnamefont {Weldeyesus}}, \emph {et~al.},\ }\bibfield
  {title} {\bibinfo {title} {Signatures of chiral superconductivity in
  rhombohedral graphene},\ }\href {https://doi.org/10.1038/s41586-025-09169-7}
  {\bibfield  {journal} {\bibinfo  {journal} {Nature}\ ,\ \bibinfo {pages} {1}}
  (\bibinfo {year} {2025})}\BibitemShut {NoStop}%
\bibitem [{\citenamefont {Pustogow}\ \emph {et~al.}(2019)\citenamefont
  {Pustogow}, \citenamefont {Luo}, \citenamefont {Chronister}, \citenamefont
  {Su}, \citenamefont {Sokolov}, \citenamefont {Jerzembeck}, \citenamefont
  {Mackenzie}, \citenamefont {Hicks}, \citenamefont {Kikugawa}, \citenamefont
  {Raghu} \emph {et~al.}}]{pustogow2019constraints}%
  \BibitemOpen
  \bibfield  {author} {\bibinfo {author} {\bibfnamefont {A.}~\bibnamefont
  {Pustogow}}, \bibinfo {author} {\bibfnamefont {Y.}~\bibnamefont {Luo}},
  \bibinfo {author} {\bibfnamefont {A.}~\bibnamefont {Chronister}}, \bibinfo
  {author} {\bibfnamefont {Y.-S.}\ \bibnamefont {Su}}, \bibinfo {author}
  {\bibfnamefont {D.}~\bibnamefont {Sokolov}}, \bibinfo {author} {\bibfnamefont
  {F.}~\bibnamefont {Jerzembeck}}, \bibinfo {author} {\bibfnamefont {A.~P.}\
  \bibnamefont {Mackenzie}}, \bibinfo {author} {\bibfnamefont {C.~W.}\
  \bibnamefont {Hicks}}, \bibinfo {author} {\bibfnamefont {N.}~\bibnamefont
  {Kikugawa}}, \bibinfo {author} {\bibfnamefont {S.}~\bibnamefont {Raghu}},
  \emph {et~al.},\ }\bibfield  {title} {\bibinfo {title} {Constraints on the
  superconducting order parameter in sr2ruo4 from oxygen-17 nuclear magnetic
  resonance},\ }\href {https://doi.org/10.1038/s41586-019-1596-2} {\bibfield
  {journal} {\bibinfo  {journal} {Nature}\ }\textbf {\bibinfo {volume} {574}},\
  \bibinfo {pages} {72} (\bibinfo {year} {2019})}\BibitemShut {NoStop}%
\bibitem [{\citenamefont {Kayyalha}\ \emph {et~al.}(2020)\citenamefont
  {Kayyalha}, \citenamefont {Xiao}, \citenamefont {Zhang}, \citenamefont
  {Shin}, \citenamefont {Jiang}, \citenamefont {Wang}, \citenamefont {Zhao},
  \citenamefont {Xiao}, \citenamefont {Zhang}, \citenamefont {Fijalkowski}
  \emph {et~al.}}]{kayyalha2020absence}%
  \BibitemOpen
  \bibfield  {author} {\bibinfo {author} {\bibfnamefont {M.}~\bibnamefont
  {Kayyalha}}, \bibinfo {author} {\bibfnamefont {D.}~\bibnamefont {Xiao}},
  \bibinfo {author} {\bibfnamefont {R.}~\bibnamefont {Zhang}}, \bibinfo
  {author} {\bibfnamefont {J.}~\bibnamefont {Shin}}, \bibinfo {author}
  {\bibfnamefont {J.}~\bibnamefont {Jiang}}, \bibinfo {author} {\bibfnamefont
  {F.}~\bibnamefont {Wang}}, \bibinfo {author} {\bibfnamefont {Y.-F.}\
  \bibnamefont {Zhao}}, \bibinfo {author} {\bibfnamefont {R.}~\bibnamefont
  {Xiao}}, \bibinfo {author} {\bibfnamefont {L.}~\bibnamefont {Zhang}},
  \bibinfo {author} {\bibfnamefont {K.~M.}\ \bibnamefont {Fijalkowski}}, \emph
  {et~al.},\ }\bibfield  {title} {\bibinfo {title} {Absence of evidence for
  chiral majorana modes in quantum anomalous hall-superconductor devices},\
  }\href {https://www.science.org/doi/10.1126/science.aax6361} {\bibfield
  {journal} {\bibinfo  {journal} {Science}\ }\textbf {\bibinfo {volume}
  {367}},\ \bibinfo {pages} {64} (\bibinfo {year} {2020})}\BibitemShut
  {NoStop}%
\bibitem [{\citenamefont {Balatsky}\ \emph {et~al.}(2006)\citenamefont
  {Balatsky}, \citenamefont {Vekhter},\ and\ \citenamefont
  {Zhu}}]{balatsky2006RMP}%
  \BibitemOpen
  \bibfield  {author} {\bibinfo {author} {\bibfnamefont {A.~V.}\ \bibnamefont
  {Balatsky}}, \bibinfo {author} {\bibfnamefont {I.}~\bibnamefont {Vekhter}},\
  and\ \bibinfo {author} {\bibfnamefont {J.-X.}\ \bibnamefont {Zhu}},\
  }\bibfield  {title} {\bibinfo {title} {Impurity-induced states in
  conventional and unconventional superconductors},\ }\href
  {https://doi.org/10.1103/RevModPhys.78.373} {\bibfield  {journal} {\bibinfo
  {journal} {Rev. Mod. Phys.}\ }\textbf {\bibinfo {volume} {78}},\ \bibinfo
  {pages} {373} (\bibinfo {year} {2006})}\BibitemShut {NoStop}%
\bibitem [{\citenamefont {L\"othman}\ and\ \citenamefont
  {Black-Schaffer}(2014)}]{lothman2014defects}%
  \BibitemOpen
  \bibfield  {author} {\bibinfo {author} {\bibfnamefont {T.}~\bibnamefont
  {L\"othman}}\ and\ \bibinfo {author} {\bibfnamefont {A.~M.}\ \bibnamefont
  {Black-Schaffer}},\ }\bibfield  {title} {\bibinfo {title} {Defects in the
  $d+id$-wave superconducting state in heavily doped graphene},\ }\href
  {https://doi.org/10.1103/PhysRevB.90.224504} {\bibfield  {journal} {\bibinfo
  {journal} {Phys. Rev. B}\ }\textbf {\bibinfo {volume} {90}},\ \bibinfo
  {pages} {224504} (\bibinfo {year} {2014})}\BibitemShut {NoStop}%
\bibitem [{\citenamefont {Ndengeyintwali}\ \emph {et~al.}(2024)\citenamefont
  {Ndengeyintwali}, \citenamefont {Heidari}, \citenamefont {Youmans},
  \citenamefont {Hosur},\ and\ \citenamefont {Ghaemi}}]{pouyan2024YSR}%
  \BibitemOpen
  \bibfield  {author} {\bibinfo {author} {\bibfnamefont {D.}~\bibnamefont
  {Ndengeyintwali}}, \bibinfo {author} {\bibfnamefont {S.}~\bibnamefont
  {Heidari}}, \bibinfo {author} {\bibfnamefont {C.}~\bibnamefont {Youmans}},
  \bibinfo {author} {\bibfnamefont {P.}~\bibnamefont {Hosur}},\ and\ \bibinfo
  {author} {\bibfnamefont {P.}~\bibnamefont {Ghaemi}},\ }\bibfield  {title}
  {\bibinfo {title} {Anomalous yu-shiba-rusinov spectrum and superconductivity
  induced magnetic interactions in materials with a topological band
  inversion},\ }\href {https://doi.org/10.1103/PhysRevB.110.085134} {\bibfield
  {journal} {\bibinfo  {journal} {Phys. Rev. B}\ }\textbf {\bibinfo {volume}
  {110}},\ \bibinfo {pages} {085134} (\bibinfo {year} {2024})}\BibitemShut
  {NoStop}%
\bibitem [{\citenamefont {Wu}\ \emph {et~al.}(2025)\citenamefont {Wu},
  \citenamefont {Hao}, \citenamefont {Chen}, \citenamefont {Cai}, \citenamefont
  {Wu}, \citenamefont {Chen}, \citenamefont {Wang}, \citenamefont {Ming},
  \citenamefont {Johnston}, \citenamefont {Zhang},\ and\ \citenamefont
  {Weitering}}]{wu2025Sn}%
  \BibitemOpen
  \bibfield  {author} {\bibinfo {author} {\bibfnamefont {X.}~\bibnamefont
  {Wu}}, \bibinfo {author} {\bibfnamefont {X.}~\bibnamefont {Hao}}, \bibinfo
  {author} {\bibfnamefont {Z.}~\bibnamefont {Chen}}, \bibinfo {author}
  {\bibfnamefont {Y.}~\bibnamefont {Cai}}, \bibinfo {author} {\bibfnamefont
  {M.}~\bibnamefont {Wu}}, \bibinfo {author} {\bibfnamefont {C.}~\bibnamefont
  {Chen}}, \bibinfo {author} {\bibfnamefont {K.}~\bibnamefont {Wang}}, \bibinfo
  {author} {\bibfnamefont {F.}~\bibnamefont {Ming}}, \bibinfo {author}
  {\bibfnamefont {S.}~\bibnamefont {Johnston}}, \bibinfo {author}
  {\bibfnamefont {R.-X.}\ \bibnamefont {Zhang}},\ and\ \bibinfo {author}
  {\bibfnamefont {H.~H.}\ \bibnamefont {Weitering}},\ }\bibfield  {title}
  {\bibinfo {title} {Microscopic fingerprint of chiral superconductivity},\
  }\href@noop {} {\bibfield  {journal} {\bibinfo  {journal} {submitted}\ }
  (\bibinfo {year} {2025})}\BibitemShut {NoStop}%
\bibitem [{\citenamefont {Pientka}\ \emph {et~al.}(2013)\citenamefont
  {Pientka}, \citenamefont {Glazman},\ and\ \citenamefont {von
  Oppen}}]{pientka2013topo}%
  \BibitemOpen
  \bibfield  {author} {\bibinfo {author} {\bibfnamefont {F.}~\bibnamefont
  {Pientka}}, \bibinfo {author} {\bibfnamefont {L.~I.}\ \bibnamefont
  {Glazman}},\ and\ \bibinfo {author} {\bibfnamefont {F.}~\bibnamefont {von
  Oppen}},\ }\bibfield  {title} {\bibinfo {title} {Topological superconducting
  phase in helical shiba chains},\ }\href
  {https://doi.org/10.1103/PhysRevB.88.155420} {\bibfield  {journal} {\bibinfo
  {journal} {Phys. Rev. B}\ }\textbf {\bibinfo {volume} {88}},\ \bibinfo
  {pages} {155420} (\bibinfo {year} {2013})}\BibitemShut {NoStop}%
\bibitem [{sup()}]{supp}%
  \BibitemOpen
  \bibinfo {note} {See the Supplemental Material for detailed
  information.}\BibitemShut {Stop}%
\bibitem [{\citenamefont {Wang}\ and\ \citenamefont
  {Wang}(2004)}]{wang2004impurity}%
  \BibitemOpen
  \bibfield  {author} {\bibinfo {author} {\bibfnamefont {Q.-H.}\ \bibnamefont
  {Wang}}\ and\ \bibinfo {author} {\bibfnamefont {Z.~D.}\ \bibnamefont
  {Wang}},\ }\bibfield  {title} {\bibinfo {title} {Impurity and interface bound
  states in ${d}_{{x}^{2}\ensuremath{-}{y}^{2}}{+id}_{\mathrm{xy}}$ and
  ${p}_{x}{+ip}_{y}$ superconductors},\ }\href
  {https://doi.org/10.1103/PhysRevB.69.092502} {\bibfield  {journal} {\bibinfo
  {journal} {Phys. Rev. B}\ }\textbf {\bibinfo {volume} {69}},\ \bibinfo
  {pages} {092502} (\bibinfo {year} {2004})}\BibitemShut {NoStop}%
\bibitem [{\citenamefont {Kaladzhyan}\ \emph {et~al.}(2016)\citenamefont
  {Kaladzhyan}, \citenamefont {Bena},\ and\ \citenamefont
  {Simon}}]{kaladzhyan2016asymptotic}%
  \BibitemOpen
  \bibfield  {author} {\bibinfo {author} {\bibfnamefont {V.}~\bibnamefont
  {Kaladzhyan}}, \bibinfo {author} {\bibfnamefont {C.}~\bibnamefont {Bena}},\
  and\ \bibinfo {author} {\bibfnamefont {P.}~\bibnamefont {Simon}},\ }\bibfield
   {title} {\bibinfo {title} {Asymptotic behavior of impurity-induced bound
  states in low-dimensional topological superconductors},\ }\href
  {https://iopscience.iop.org/article/10.1088/0953-8984/28/48/485701}
  {\bibfield  {journal} {\bibinfo  {journal} {Journal of Physics: Condensed
  Matter}\ }\textbf {\bibinfo {volume} {28}},\ \bibinfo {pages} {485701}
  (\bibinfo {year} {2016})}\BibitemShut {NoStop}%
\bibitem [{\citenamefont {Ming}\ \emph {et~al.}(2023)\citenamefont {Ming},
  \citenamefont {Wu}, \citenamefont {Chen}, \citenamefont {Wang}, \citenamefont
  {Mai}, \citenamefont {Maier}, \citenamefont {Strockoz}, \citenamefont
  {Venderbos}, \citenamefont {Gonz{\'a}lez}, \citenamefont {Ortega} \emph
  {et~al.}}]{ming2023evidence}%
  \BibitemOpen
  \bibfield  {author} {\bibinfo {author} {\bibfnamefont {F.}~\bibnamefont
  {Ming}}, \bibinfo {author} {\bibfnamefont {X.}~\bibnamefont {Wu}}, \bibinfo
  {author} {\bibfnamefont {C.}~\bibnamefont {Chen}}, \bibinfo {author}
  {\bibfnamefont {K.~D.}\ \bibnamefont {Wang}}, \bibinfo {author}
  {\bibfnamefont {P.}~\bibnamefont {Mai}}, \bibinfo {author} {\bibfnamefont
  {T.~A.}\ \bibnamefont {Maier}}, \bibinfo {author} {\bibfnamefont
  {J.}~\bibnamefont {Strockoz}}, \bibinfo {author} {\bibfnamefont
  {J.}~\bibnamefont {Venderbos}}, \bibinfo {author} {\bibfnamefont
  {C.}~\bibnamefont {Gonz{\'a}lez}}, \bibinfo {author} {\bibfnamefont
  {J.}~\bibnamefont {Ortega}}, \emph {et~al.},\ }\bibfield  {title} {\bibinfo
  {title} {Evidence for chiral superconductivity on a silicon surface},\ }\href
  {https://doi.org/10.1038/s41567-022-01889-1} {\bibfield  {journal} {\bibinfo
  {journal} {Nature Physics}\ }\textbf {\bibinfo {volume} {19}},\ \bibinfo
  {pages} {500} (\bibinfo {year} {2023})}\BibitemShut {NoStop}%
\bibitem [{\citenamefont {Marchetti}\ \emph {et~al.}(2025)\citenamefont
  {Marchetti}, \citenamefont {Bunney}, \citenamefont {Di~Sante},\ and\
  \citenamefont {Rachel}}]{rachel2025electronic}%
  \BibitemOpen
  \bibfield  {author} {\bibinfo {author} {\bibfnamefont {L.}~\bibnamefont
  {Marchetti}}, \bibinfo {author} {\bibfnamefont {M.}~\bibnamefont {Bunney}},
  \bibinfo {author} {\bibfnamefont {D.}~\bibnamefont {Di~Sante}},\ and\
  \bibinfo {author} {\bibfnamefont {S.}~\bibnamefont {Rachel}},\ }\bibfield
  {title} {\bibinfo {title} {Electronic structure, spin-orbit interaction, and
  electron-phonon coupling of triangular adatom lattices on semiconductor
  substrates},\ }\href {https://doi.org/10.1103/PhysRevB.111.125115} {\bibfield
   {journal} {\bibinfo  {journal} {Phys. Rev. B}\ }\textbf {\bibinfo {volume}
  {111}},\ \bibinfo {pages} {125115} (\bibinfo {year} {2025})}\BibitemShut
  {NoStop}%
\bibitem [{\citenamefont {Bernevig}\ \emph {et~al.}(2006)\citenamefont
  {Bernevig}, \citenamefont {Hughes},\ and\ \citenamefont
  {Zhang}}]{bernevig2006quantum}%
  \BibitemOpen
  \bibfield  {author} {\bibinfo {author} {\bibfnamefont {B.~A.}\ \bibnamefont
  {Bernevig}}, \bibinfo {author} {\bibfnamefont {T.~L.}\ \bibnamefont
  {Hughes}},\ and\ \bibinfo {author} {\bibfnamefont {S.-C.}\ \bibnamefont
  {Zhang}},\ }\bibfield  {title} {\bibinfo {title} {Quantum spin hall effect
  and topological phase transition in hgte quantum wells},\ }\href
  {https://www.science.org/doi/10.1126/science.1133734} {\bibfield  {journal}
  {\bibinfo  {journal} {science}\ }\textbf {\bibinfo {volume} {314}},\ \bibinfo
  {pages} {1757} (\bibinfo {year} {2006})}\BibitemShut {NoStop}%
\bibitem [{\citenamefont {Asahi}\ and\ \citenamefont
  {Nagaosa}(2012)}]{asahi2012defect}%
  \BibitemOpen
  \bibfield  {author} {\bibinfo {author} {\bibfnamefont {D.}~\bibnamefont
  {Asahi}}\ and\ \bibinfo {author} {\bibfnamefont {N.}~\bibnamefont
  {Nagaosa}},\ }\bibfield  {title} {\bibinfo {title} {Topological indices,
  defects, and majorana fermions in chiral superconductors},\ }\href
  {https://doi.org/10.1103/PhysRevB.86.100504} {\bibfield  {journal} {\bibinfo
  {journal} {Phys. Rev. B}\ }\textbf {\bibinfo {volume} {86}},\ \bibinfo
  {pages} {100504} (\bibinfo {year} {2012})}\BibitemShut {NoStop}%
\bibitem [{\citenamefont {Hughes}\ \emph {et~al.}(2014)\citenamefont {Hughes},
  \citenamefont {Yao},\ and\ \citenamefont {Qi}}]{hughes2014dislocation}%
  \BibitemOpen
  \bibfield  {author} {\bibinfo {author} {\bibfnamefont {T.~L.}\ \bibnamefont
  {Hughes}}, \bibinfo {author} {\bibfnamefont {H.}~\bibnamefont {Yao}},\ and\
  \bibinfo {author} {\bibfnamefont {X.-L.}\ \bibnamefont {Qi}},\ }\bibfield
  {title} {\bibinfo {title} {Majorana zero modes in dislocations of
  ${\mathrm{sr}}_{2}{\mathrm{ruo}}_{4}$},\ }\href
  {https://doi.org/10.1103/PhysRevB.90.235123} {\bibfield  {journal} {\bibinfo
  {journal} {Phys. Rev. B}\ }\textbf {\bibinfo {volume} {90}},\ \bibinfo
  {pages} {235123} (\bibinfo {year} {2014})}\BibitemShut {NoStop}%
\bibitem [{\citenamefont {Hu}\ and\ \citenamefont
  {Zhang}(2024)}]{hu2024dislocation}%
  \BibitemOpen
  \bibfield  {author} {\bibinfo {author} {\bibfnamefont {L.-H.}\ \bibnamefont
  {Hu}}\ and\ \bibinfo {author} {\bibfnamefont {R.-X.}\ \bibnamefont {Zhang}},\
  }\bibfield  {title} {\bibinfo {title} {Dislocation majorana bound states in
  iron-based superconductors},\ }\href
  {https://doi.org/10.1038/s41467-024-46618-9} {\bibfield  {journal} {\bibinfo
  {journal} {Nature Communications}\ }\textbf {\bibinfo {volume} {15}},\
  \bibinfo {pages} {2337} (\bibinfo {year} {2024})}\BibitemShut {NoStop}%
\bibitem [{\citenamefont {Brei\o{}}\ \emph {et~al.}(2024)\citenamefont
  {Brei\o{}}, \citenamefont {Kreisel}, \citenamefont {Roig}, \citenamefont
  {Hirschfeld},\ and\ \citenamefont {Andersen}}]{breio2024dislocation}%
  \BibitemOpen
  \bibfield  {author} {\bibinfo {author} {\bibfnamefont {C.~N.}\ \bibnamefont
  {Brei\o{}}}, \bibinfo {author} {\bibfnamefont {A.}~\bibnamefont {Kreisel}},
  \bibinfo {author} {\bibfnamefont {M.}~\bibnamefont {Roig}}, \bibinfo {author}
  {\bibfnamefont {P.~J.}\ \bibnamefont {Hirschfeld}},\ and\ \bibinfo {author}
  {\bibfnamefont {B.~M.}\ \bibnamefont {Andersen}},\ }\bibfield  {title}
  {\bibinfo {title} {Time-reversal symmetry breaking from lattice dislocations
  in superconductors},\ }\href {https://doi.org/10.1103/PhysRevB.109.014505}
  {\bibfield  {journal} {\bibinfo  {journal} {Phys. Rev. B}\ }\textbf {\bibinfo
  {volume} {109}},\ \bibinfo {pages} {014505} (\bibinfo {year}
  {2024})}\BibitemShut {NoStop}%
\bibitem [{\citenamefont {Fu}\ and\ \citenamefont {Kane}(2008)}]{fu2008vortex}%
  \BibitemOpen
  \bibfield  {author} {\bibinfo {author} {\bibfnamefont {L.}~\bibnamefont
  {Fu}}\ and\ \bibinfo {author} {\bibfnamefont {C.~L.}\ \bibnamefont {Kane}},\
  }\bibfield  {title} {\bibinfo {title} {Superconducting proximity effect and
  majorana fermions at the surface of a topological insulator},\ }\href
  {https://doi.org/10.1103/PhysRevLett.100.096407} {\bibfield  {journal}
  {\bibinfo  {journal} {Phys. Rev. Lett.}\ }\textbf {\bibinfo {volume} {100}},\
  \bibinfo {pages} {096407} (\bibinfo {year} {2008})}\BibitemShut {NoStop}%
\bibitem [{\citenamefont {Hosur}\ \emph {et~al.}(2011)\citenamefont {Hosur},
  \citenamefont {Ghaemi}, \citenamefont {Mong},\ and\ \citenamefont
  {Vishwanath}}]{hosur2011vortex}%
  \BibitemOpen
  \bibfield  {author} {\bibinfo {author} {\bibfnamefont {P.}~\bibnamefont
  {Hosur}}, \bibinfo {author} {\bibfnamefont {P.}~\bibnamefont {Ghaemi}},
  \bibinfo {author} {\bibfnamefont {R.~S.~K.}\ \bibnamefont {Mong}},\ and\
  \bibinfo {author} {\bibfnamefont {A.}~\bibnamefont {Vishwanath}},\ }\bibfield
   {title} {\bibinfo {title} {Majorana modes at the ends of superconductor
  vortices in doped topological insulators},\ }\href
  {https://doi.org/10.1103/PhysRevLett.107.097001} {\bibfield  {journal}
  {\bibinfo  {journal} {Phys. Rev. Lett.}\ }\textbf {\bibinfo {volume} {107}},\
  \bibinfo {pages} {097001} (\bibinfo {year} {2011})}\BibitemShut {NoStop}%
\bibitem [{\citenamefont {Hu}\ \emph {et~al.}(2022)\citenamefont {Hu},
  \citenamefont {Wu}, \citenamefont {Liu},\ and\ \citenamefont
  {Zhang}}]{hu2022vortex}%
  \BibitemOpen
  \bibfield  {author} {\bibinfo {author} {\bibfnamefont {L.-H.}\ \bibnamefont
  {Hu}}, \bibinfo {author} {\bibfnamefont {X.}~\bibnamefont {Wu}}, \bibinfo
  {author} {\bibfnamefont {C.-X.}\ \bibnamefont {Liu}},\ and\ \bibinfo {author}
  {\bibfnamefont {R.-X.}\ \bibnamefont {Zhang}},\ }\bibfield  {title} {\bibinfo
  {title} {Competing vortex topologies in iron-based superconductors},\ }\href
  {https://doi.org/10.1103/PhysRevLett.129.277001} {\bibfield  {journal}
  {\bibinfo  {journal} {Phys. Rev. Lett.}\ }\textbf {\bibinfo {volume} {129}},\
  \bibinfo {pages} {277001} (\bibinfo {year} {2022})}\BibitemShut {NoStop}%
\bibitem [{\citenamefont {Hu}\ and\ \citenamefont
  {Zhang}(2023)}]{hu2023topological}%
  \BibitemOpen
  \bibfield  {author} {\bibinfo {author} {\bibfnamefont {L.-H.}\ \bibnamefont
  {Hu}}\ and\ \bibinfo {author} {\bibfnamefont {R.-X.}\ \bibnamefont {Zhang}},\
  }\bibfield  {title} {\bibinfo {title} {Topological superconducting vortex
  from trivial electronic bands},\ }\href
  {https://doi.org/10.1038/s41467-023-36347-w} {\bibfield  {journal} {\bibinfo
  {journal} {Nature Communications}\ }\textbf {\bibinfo {volume} {14}},\
  \bibinfo {pages} {640} (\bibinfo {year} {2023})}\BibitemShut {NoStop}%
\bibitem [{\citenamefont {Holmvall}\ and\ \citenamefont
  {Black-Schaffer}(2023)}]{holmvall2023coreless}%
  \BibitemOpen
  \bibfield  {author} {\bibinfo {author} {\bibfnamefont {P.}~\bibnamefont
  {Holmvall}}\ and\ \bibinfo {author} {\bibfnamefont {A.~M.}\ \bibnamefont
  {Black-Schaffer}},\ }\bibfield  {title} {\bibinfo {title} {Coreless vortices
  as direct signature of chiral $d$-wave superconductivity},\ }\href
  {https://doi.org/10.1103/PhysRevB.108.L100506} {\bibfield  {journal}
  {\bibinfo  {journal} {Phys. Rev. B}\ }\textbf {\bibinfo {volume} {108}},\
  \bibinfo {pages} {L100506} (\bibinfo {year} {2023})}\BibitemShut {NoStop}%
\bibitem [{\citenamefont {Holmvall}\ \emph {et~al.}(2023)\citenamefont
  {Holmvall}, \citenamefont {Wall-Wennerdal},\ and\ \citenamefont
  {Black-Schaffer}}]{holmvall2023robust}%
  \BibitemOpen
  \bibfield  {author} {\bibinfo {author} {\bibfnamefont {P.}~\bibnamefont
  {Holmvall}}, \bibinfo {author} {\bibfnamefont {N.}~\bibnamefont
  {Wall-Wennerdal}},\ and\ \bibinfo {author} {\bibfnamefont {A.~M.}\
  \bibnamefont {Black-Schaffer}},\ }\bibfield  {title} {\bibinfo {title}
  {Robust and tunable coreless vortices and fractional vortices in chiral
  $d$-wave superconductors},\ }\href
  {https://doi.org/10.1103/PhysRevB.108.094511} {\bibfield  {journal} {\bibinfo
   {journal} {Phys. Rev. B}\ }\textbf {\bibinfo {volume} {108}},\ \bibinfo
  {pages} {094511} (\bibinfo {year} {2023})}\BibitemShut {NoStop}%
\bibitem [{\citenamefont {Mermin}(1979)}]{mermin1979RMP}%
  \BibitemOpen
  \bibfield  {author} {\bibinfo {author} {\bibfnamefont {N.~D.}\ \bibnamefont
  {Mermin}},\ }\bibfield  {title} {\bibinfo {title} {The topological theory of
  defects in ordered media},\ }\href
  {https://doi.org/10.1103/RevModPhys.51.591} {\bibfield  {journal} {\bibinfo
  {journal} {Rev. Mod. Phys.}\ }\textbf {\bibinfo {volume} {51}},\ \bibinfo
  {pages} {591} (\bibinfo {year} {1979})}\BibitemShut {NoStop}%
\end{thebibliography}%

\onecolumngrid

\appendix

\subsection{\large{Supplemental Material for ``Deciphering Chiral Superconductivity via Impurity Bound States"}}

\section{Classification of Chiral Pairing Orders}
\label{sec:pairing class}

In this section, we classify the pairing matrix $\Delta({\bf k})$ for the 2d Bogoliubov–de Gennes (BdG) Hamiltonian described by Eq.~1 of the main text, based on its property under an $n$-fold rotational symmetry $C_n$. Let us define the rotation operator $C_n$ of the normal state as $C_n=\exp({i\frac{2\pi s}{n}\sigma_z})$, where $s$ denotes the half-integer-valued electron spin. In spin-orbital coupled systems, $s$ should be viewed as the $\hat{z}$-component angular momentum of electrons. A general Cooper pairing fulfilling both $C_n$ symmetry and Fermi statistics must satisfy:
\begin{equation}
\label{eq.Cn and Fermi sta}
    \begin{split}
        C_n\Delta({\bf k})C_n^T & =e^{-i\frac{2\pi j_z}{n}}\Delta(C_n{\bf k}),\\
        \Delta({\bf k}) & =-\Delta^T(-{\bf k}),
    \end{split}
\end{equation}
where $j_z\in\mathbb{Z}_n$ is the angular momentum of Cooper pairs. Under the 2 by 2 spin basis, any pairing order can be expanded with the spin Pauli matrices as
\begin{equation}
    \Delta({\bf k})=d_0({\bf k}) \sigma_0+\Sigma_{i=x,y,z}d_i({\bf k}) \sigma_i.
\end{equation}
Applying the constraints in Eq.~\ref{eq.Cn and Fermi sta}, we have
\begin{equation}
    \begin{split}
        & e^{i\frac{2\pi s}{n}\sigma_z}\Delta({\bf k})e^{i\frac{2\pi s}{n}\sigma_z}  =e^{-i\frac{2\pi j_z}{n}}\Delta(C_n{\bf k})\\
       & d_{0,x,z}({\bf k})  =-d_{0,x,z}({-\bf k}),\ \ d_y({\bf k}) =d_y({-\bf k}).
    \end{split}
\end{equation}
In the following, we will discuss the possible chiral pairing orders for each individual $C_n$, which will eventually lead to Table.~I.

\subsection{$n=2$}
For electrons, we have $C_2=e^{i\frac{2\pi s}{2}\sigma_z}=(-1)^{s}\sigma_z$ with $C_2{\bf k}=-{\bf k}$. Eq.~\ref{eq.Cn and Fermi sta} now becomes: 
\begin{eqnarray}
        && d_{0,z}({\bf k}) = (-1)^{-j_z-2s}d_{0,z} (-{\bf k}),\ 
        d_{x,y}({\bf k})=(-1)^{-j_z-2s-1}d_{x,y}(-{\bf k}),\nonumber \\
        && d_{0,x,z}({\bf k})  =-d_{0,x,z}({-\bf k}),\ \ d_y({\bf k}) =d_y({-\bf k}).
\end{eqnarray}
Since $2s\in\mathbb{Z}_{odd}$, we have $(-1)^{2s}=-1$. Plugging in Eq.~\ref{eq.Cn and Fermi sta}, we get:
\begin{equation}
    \begin{split}
        \begin{cases}
            d_{0,z}({\bf k})=-d_{0,z}(-{\bf k}),\ d_x({\bf k})=d_x(-{\bf k})=-d_x(-{\bf k}),\ d_y({\bf k})=d_y(-{\bf k}), & \text{if }j_z=0;\\
            d_{0,z}({\bf k})=-d_{0,z}(-{\bf k})=d_{0,z}(-{\bf k}),\ d_x({\bf k})=-d_x(-{\bf k}),\ d_y({\bf k})=-d_y(-{\bf k})=d_y(-{\bf k}), & \text{if }j_z=1.
        \end{cases}
    \end{split}
\end{equation}
Consider expanding $d_i({\bf k})$ up to ${\cal O}(k^2)$, we immediately have:
\begin{itemize}
    \item $j_z=0$: $i\sigma_y, k_\pm\sigma_{0,z}, k_\pm^2\sigma_y$.
    \item $j_z=1$: $k_\pm\sigma_{x}$.
\end{itemize}
Notably, $k_\pm \sigma_{0,z}$ represents chiral pairing order with a nontrivial winding number of $l=\pm 1$, but with a vanishing $j_z$.

\subsection{$n>2$}
We now extend the above discussion to $n=3,4,6$, where $C_n=e^{i\frac{2\pi s}{n}\sigma_z}$. We generally have
\begin{equation}
    \begin{bmatrix}
        (d_0({\bf k})+d_z({\bf k}))e^{i\frac{4\pi s}{n}} & d_x({\bf k})-id_y({\bf k})\\
        d_x({\bf k})+id_y({\bf k}) & (d_0({\bf k})-d_z({\bf k}))e^{-i\frac{4\pi s}{n}}
    \end{bmatrix}  =e^{-i\frac{2\pi j_z}{n}}\begin{bmatrix}
         d_0(C_n{\bf k})+d_z(C_n{\bf k}) & d_x(C_n{\bf k})-id_y(C_n{\bf k})\\
        d_x(C_n{\bf k})+id_y(C_n{\bf k}) & d_0(C_n{\bf k})-d_z(C_n{\bf k})
    \end{bmatrix},
\end{equation}
leading to,
\begin{equation}
\label{eq. sim constraints}
\begin{split}
    d_0({\bf k})\pm d_z({\bf k})=& e^{i\frac{2\pi}{n}(-j_z \mp 2s)}(d_{0,z}(C_n{\bf k})\pm d_z(C_n{\bf k}))=d_0(-{\bf k})\pm d_z(-{\bf k}),\\
    d_x({\bf k})=& e^{-i\frac{2\pi j_z}{n}}d_x(C_n{\bf k})=-d_x(-{\bf k}),\\
    d_y({\bf k})=& e^{-i\frac{2\pi j_z}{n}}d_y(C_n{\bf k})=d_y(-{\bf k}).
\end{split}
\end{equation}
From Eq.~\ref{eq. sim constraints}, we find that the pairing matrices $\sigma_0 \pm \sigma_z$ contribute angular momentum $\pm 2s$ to the total Cooper-pair angular momentum $j_z$, while $\sigma_x$ and $\sigma_y$ contribute zero. The total $j_z$ arises from the sum of this matrix contribution and that of the form factor $d_i(\mathbf{k})$. Importantly, for $n > 2$, $C_n$ symmetry requires the pairing to involve a superposition of $\sigma_0$ and $\sigma_z$, which nevertheless only gaps out one of the spin-degenerate Fermi surfaces. This constraint marks a key distinction from the $C_2$-symmetric case. To achieve fully gapped chiral superconductivity, we therefore focus on form factors of the form $d_{x,y}(\mathbf{k}) \sim (k_\pm)^p$ with integer $p$. Fermi statistics further require $p$ to be odd for $d_x(\mathbf{k})$ and even for $d_y(\mathbf{k})$.

Now it is straightforward to show that
\begin{itemize}
    \item $i\sigma_y$: The $s$-wave spin-singlet pairing always has $j_z=0$.
    \item $k_\pm^p\sigma_x$: $p$ is odd. The corresponding $j_z\equiv \pm p$ (mod $n$) for $k_\pm^p$, respectively. Note that when $n=3$, a chiral $f\pm if$ pairing has $j_z=0$.
    \item $k_\pm^p\sigma_y$: $p$ is even. The corresponding $j_z\equiv \pm p$ (mod $n$) for $k_\pm^p$, respectively. When $n=4$, a chiral $g$-wave pairing with $p=4$ has $j_z=0$. 
\end{itemize}

\section{Analytical Proof of Nodal condition}
\label{sec:analytical proof}

In this section, we provide an analytical proof for the nodal condition in Eqs.~3 and 5 of the main text, which is the central result of this work. Our proof should hold for any general 2d $C_n$-invariant superconductor with a pair of spin-degenerate Fermi surfaces. Let us start by briefly reviewing our setup. The $k$-space BdG Hamiltonian is
\begin{equation}
    H({\bf k})=\begin{bmatrix}
        \xi_{\bf k} & \Delta(\bf k)\\
        \Delta^\dagger({\bf k}) & -\xi_{-\bf k}
    \end{bmatrix},
    \label{eq:BdG-spin-degenerate}
\end{equation}
where $\xi_{\bf k}=E({\bf k})-\mu$. The $n$-fold rotation symmetry $C_n$ requires the electron dispersion to follow $E(k,\theta)=E(k,\theta+2\pi/n)$, where $(k,\theta)$ are polar coordinates in momentum space.

As shown in Sec.~\ref{sec:pairing class}, the pairing order can always be written as $\Delta({\bf k})=f({\bf k})M$, a form factor part $f({\bf k})\in\mathbb{C}$ and a matrix part $M\in\sigma_{0,x,y,z}$. Crucially, the form factor must make an {\it integer-valued} contribution to $j_z$, the angular momentum of the Cooper pairs. This is equivalent to  
\begin{equation}
    f(k,\theta+\frac{2\pi}{n})=e^{i\frac{2\pi l}{n}}f(k,\theta),\ \ l\in\mathbb{Z}.
    \label{eq:form-factor-Cn}
\end{equation}
Physically, $l$ is exactly the chiral winding number of the pairing order around the Fermi surfaces. Note that the matrix part $M$ may also contribute to $j_z$ if it transforms nontrivially under $C_n$, e.g.,  $C_nMC_n^T=e^{i\Tilde{l}}M$, which will lead to a net pairing angular momentum of $j_z=l+\Tilde{l}$. \\

We now consider a point-like impurity potential $V\delta({\bf r}={\bf 0})$ located at the origin. The bound-state energy $E$ and corresponding wavefunction $\psi({\bf r}={\bf 0})$ yield~\cite{pientka2013topo}:
\begin{equation}
    (\mathbbm{1}_4-V\cdot\mathcal{G}(E,{\bf r}={\bf 0}))\psi({\bf r}={\bf 0})=0.
\end{equation}
where the real-space Green's function is given by:
\begin{eqnarray}
    && \mathcal{G}(E,{\bf r}={\bf 0})=\int\frac{d^2k}{(2\pi)^2} G(E,{\bf k}),\nonumber \\
    && G(E,{\bf k})=\frac{1}{E+i0^+-H({\bf k})}.
\end{eqnarray}

The key to deriving the nodal condition is to identify when the real-space anomalous propagator ${\cal F}(E,{\bf r}=0)$ vanishes at the defect center. For convenience, let us define the following notations:
\begin{eqnarray}
    \Omega_\pm({\bf k}) \equiv E+i0^+\mp \xi_{\pm\bf k}, \quad {\cal S} \equiv \Omega_+({\bf k})-\Delta({\bf k})\Omega_-^{-1}({\bf k})\Delta^\dagger({\bf k}),
\end{eqnarray}
where ${\cal S}({\bf k})$ is the {\it Schur complement} of $\Omega_-({\bf k})$ with respect to $G(E,{\bf k})$. Applying the inversion formula for block-wise matrices, we find the $k$-space anomalous propagator to be $F(E,{\bf k})={\cal S}^{-1}({\bf k})\Delta({\bf k})\Omega_-^{-1}({\bf k})$. Since $\Omega_-({\bf k})$ is diagonal, it follows that its Schur complement is diagonal as well. Then, a general matrix element of the anomalous propagator is given by
\begin{equation}
    F_{ss'}(E,{\bf k})=[\mathcal{S}^{-1}({\bf k})]_{ss}[M]_{ss'}[\Omega_-^{-1}({\bf k})]_{s's'}f({\bf k}).
    \label{eq:anomalous-propagator-element}
\end{equation}
Back to the real-space representation, the corresponding on-site anomalous propagator is
\begin{equation}
    \mathcal{F}_{ss'}(E,{\bf r}={\bf 0}) =\int\frac{d^2k}{(2\pi)^2}F_{ss'}(E,{\bf k})=\iint \frac{kdkd\theta}{(2\pi)^2}F_{ss'}(E,k,\theta).
    \label{eq:anomalous-propagator-onsite}
\end{equation}
\\

Now we proceed to explore how $C_n$ constrains $\mathcal{F}_{ss'}(E,{\bf r}={\bf 0})$. Since the electronic dispersion relation satisfies $\xi(k,\theta) = \xi(k,\theta+ 2\pi/n)$, we find that
\begin{equation}
        [\Omega_-({k,\theta+\frac{2\pi}{n}})]_{ss} =[\Omega_-({k,\theta})]_{ss},\quad
        [\mathcal{S}(k,\theta+\frac{2\pi}{n})]_{ss} =[\mathcal{S}(k,\theta)]_{ss},
\end{equation}
where we have exploited Eq.~\ref{eq:form-factor-Cn}. Together with Eq.~\ref{eq:anomalous-propagator-element}, we hence arrive at an important relation,
\begin{eqnarray}
    F_{ss'}(E,k,\theta+\frac{2\pi}{n})=e^{i\frac{2\pi}{n}l}F_{ss'}(E,k,\theta).
\end{eqnarray}
As a consequence, we can always rewrite the angular integral of Eq.~\ref{eq:anomalous-propagator-onsite} into a sum of $n$ rotation-related sectors:
\begin{equation}
    \begin{split}
        \mathcal{F}_{ss'}(E,{\bf r}= {\bf 0})=& \sum_{m=0}^{n-1}\int \frac{kdk}{(2\pi)^2}\int_{2m\pi/n}^{2(m+1)\pi/n}d\theta \text{ }F_{ss'}(E,k,\theta)\\
    =& \sum_{m=0}^{n-1}\int \frac{kdk}{(2\pi)^2} \int_0^{2\pi/n}d\theta\ F_{ss'}(E,k,\theta+\frac{2\pi}{n}m)\\
    =& \int \frac{kdk}{(2\pi)^2} \int_0^{2\pi/n}d\theta\text{ }F_{ss'}(E,k,\theta)\sum_{m=0}^{n-1}e^{2\pi i l m/n}
    \end{split}
\end{equation}
Finally, using the geometric series:
\begin{equation}
    \sum_{m=0}^{n-1}e^{2\pi i l m/n}=\frac{1-e^{2\pi il}}{1-e^{2\pi il/n}}=\begin{cases}
n, & \text{if } l/n\in\mathbb{Z}, \\
0,   & \text{otherwise},
\end{cases}
\end{equation}
we arrive at Eq.~3 of the main text,
\begin{equation}
    \mathcal{F}_{ss'}(E,{\bf r})=\begin{cases}
n \int \frac{kdk}{(2\pi)^2} \int_0^{2\pi/n}d\theta\text{ }F_{ss'}(E,k,\theta), & \text{if } l/n\in\mathbb{Z}, \\
0,   & \text{otherwise}.
\end{cases}
\end{equation}
As discussed in the main text, when the anomalous propagator vanishes at the impurity site, the bound state at ${\bf r}=0$ must be either electron-like or hole-like. This necessarily leads to the characteristic ``node-antinode" structure in the LDOS.

\section{Rashba Effects}

The proof in Sec.~\ref{sec:analytical proof} assumes a spin-degenerate normal state. In this section, we will relax this assumption and explore the consequences of Rashba effects that lift this spin degeneracy. Let us start with a BdG Hamiltonian in the basis $\Psi({\bf k})=(c_{{\bf k},\uparrow},c_{{\bf k},\downarrow},c_{{-\bf k},\uparrow}^\dagger,c_{{-\bf k},\uparrow}^\dagger)$, given by
\begin{equation}
    H({\bf k})=\begin{bmatrix}
        h({\bf k}) & \Delta({\bf k})\\
        \Delta^\dagger({\bf k}) & -h^T(-{\bf k})
    \end{bmatrix}.
\end{equation}
The normal state considered here may include a Rashba term, and we only require it to respect a time reversal symmetry (TRS) $\Theta=i\sigma_y\mathcal{K}$, with
\begin{equation}
   \Theta h({\bf k})\Theta^{-1}=h(-{\bf k}),
\end{equation}
where $\mathcal{K}$ denotes complex conjugation.

Let $\ket{u_{n,s}({\bf k})} = (\ket{u_{n,1}({\bf k})}, \ket{u_{n,2}({\bf k})})^T$ be a Kramers pair of Bloch states that diagonalize $h(k)$, where $n$ is a band index and $s=1,2$ labels the Kramers pair. Noting that $h(-{\bf k})\Theta\ket{u_{n,s}({\bf k})}=E_{\bf k}\Theta\ket{u_{n,s}({\bf k})}$. This leads to  
\begin{equation}
    \Theta\begin{pmatrix}
        \ket{u_{n,1}({\bf k})} \\
        \ket{u_{n,2}({\bf k})} \\
    \end{pmatrix} = \begin{pmatrix}
        0 & -e^{-i\alpha_{\bf k}} \\
        e^{-i\alpha_{\bf -k}} & 0 \\
    \end{pmatrix} \begin{pmatrix}
        \ket{u_{n,1}({-\bf k})} \\
        \ket{u_{n,2}({-\bf k})} \\
    \end{pmatrix},
\end{equation}
where $\alpha_{\bf k}$ is the time-reversal-invariant gauge. Since the gauge choice does not affect the physics we will focus on below, for simplicity, we fix $\alpha_k=0$. 

\subsection{Singlet Pairing}

Now we are ready to project the ``bare" pairing function onto such a Kramers pair of bands. We first focus on a spin-singlet pairing with a form factor $f({\bf k})=f(-{\bf k})$ is given by $\Delta({\bf k})=i\sigma_yf({\bf k})$. The corresponding projected pairing is
\begin{eqnarray}
    (\widetilde{\Delta}_n)_{12} &= & \bra{u_{n,1}({\bf k})}\Delta({\bf k})\ket{u^*_{n,2}({-\bf k})}= -f({\bf k}) \bra{u_{n,1}({\bf k})}i\sigma_y \Theta\ket{u_{n,1}^*({\bf k})} = -f({\bf k}) \bra{u_{n,1}({\bf k})}(i\sigma_y )^2\ket{u_{n,1}({\bf k})} = f({\bf k}),  \nonumber \\
    (\widetilde{\Delta}_n)_{21} & = & \bra{u_{n,2}({\bf k})}\Delta({\bf k})\ket{u^*_{n,1}({-\bf k})}=  f({\bf k}) \bra{u_{n,2}({\bf k})}i\sigma_y \Theta\ket{u_{n,2}^*({\bf k})} = -f({\bf k}), \nonumber \\
    (\widetilde{\Delta}_n)_{11} &=& (\widetilde{\Delta}_n)_{22} = 0. 
\end{eqnarray}
As a result, we obtain
\begin{equation}
\widetilde{\Delta}_n = i s_y f({\bf k}),
\label{eq:singlet projection}
\end{equation}
which corresponds to a spin-singlet pairing in the pseudo-spin (band) basis, where $s_y$ is the Pauli matrix acting within the Kramers doublet. Importantly, Eq.~\ref{eq:singlet projection} holds generically for time-reversal-invariant systems, including those with Rashba spin-orbit coupling. Therefore, for spin-singlet pairing, projection onto the time-reversal-related band doublets always yields a low-energy Hamiltonian of the same form as Eq.~\ref{eq:BdG-spin-degenerate}, ensuring that the nodal condition derived earlier continues to apply.

\subsection{Triplet Pairing}

When the pairing is spin-triplet, introducing Rashba spin-orbit coupling leads to two important consequences that are absent in the singlet case. First, spin-triplet pairing often competes with the Rashba term $h_R({\bf k})$, which is reflected at the matrix level by $\{ h_R({\bf k}), \Delta({\bf k}) \} \neq 0$. This competition can prevent the system from developing a full superconducting gap, particularly when the Rashba coupling is sufficiently strong. For instance, this behavior has been noted in the context of the Sn/Si(111) system, where the combination of chiral $p$-wave triplet pairing and Rashba interaction leads to gapless quasiparticle spectra~\cite{wu2025Sn}. Second, the Rashba coupling significantly alters the spin texture along the Fermi surface. When a spin-triplet pairing is projected onto such a Rashba-twisted Fermi surface, the resulting effective pairing matrix may qualitatively differ from its original form.

As an intuitive example, we consider a chiral $p$-wave triplet pairing with $\Delta({\bf k})=\Delta_0 k_+\sigma_x$ for a pair of Rashba electron bands,
\begin{equation}
    h_R({\bf k})=k_x\sigma_y-k_y\sigma_x.
\end{equation}
Using the lattice regularization: $k_x\rightarrow{\sin{k_x}}$, $k_y\rightarrow{\sin{k_y}}$, the Hamiltonian $h_R({\bf k})$ and pairing $\Delta({\bf k})$ become:
\begin{equation}
        h({\bf k})= \sin{k_x}\sigma_y-\sin{k_y}\sigma_x, \quad
        \Delta({\bf k})= \Delta_0(\sin{k_x}+i\sin{k_y})\sigma_x.
\end{equation}
The eigenstates of $h_R({\bf k})$ are
\begin{equation}
    \psi_\pm({\bf k})=\frac{e^{i\alpha_\pm}}{\sqrt{2}}(\mp ie^{i\theta_{\bf k+}}, 1)^T,\ E_\pm=\pm E_0,
\end{equation}
where $E_0=\sqrt{\sin^2{k_x}+\sin^2{k_y}}$, $\alpha_\pm$ are gauge factors and $\theta_{\bf k\pm}=\arg{(\sin{k_x}\pm i\sin{k_y})}$. We first choose the gauge $\alpha_+=0,\ \alpha_-=ie^{\theta_{\bf k-}}$,  such that the time reversal operator $\Theta=i\sigma_y\mathcal{K}$ satisfies:
\begin{equation}
    \Theta\psi_+({\bf k})=\psi_-({\bf k}),\ \Theta\psi_-({\bf k})=-\psi_+({\bf k}).
\end{equation}

Define the unitary matrix $U({\bf k})=(\psi_+({\bf k}), \psi_-({\bf k}))$, so that $U^\dagger({\bf k})h({\bf k})U({\bf k})$ and $U^T({-\bf k})h^T({-\bf k})U^*({-\bf k})$ are diagonal. The pairing projected onto the spinor basis then reads:
\begin{equation}
    \widetilde\Delta({\bf k}) = U^\dagger({\bf k})\Delta({\bf k})U^*({-\bf k}) = (\sin{k_x}+i\sin{k_y})\begin{bmatrix}
        \sin{k_y}/E_0 & (i\sin{k_x}\sin{k_y}-\sin^2{k_x})/E_0^2\\
        (-i\sin{k_x}\sin{k_y}-\sin^2{k_x})/E_0^2 & -\sin{k_y}/E_0
    \end{bmatrix},
\end{equation}
which contains both intraband and interband components and makes it hard for us to read out the BdG band structure. 

This inspires us to consider a difference choice of gauge with $\alpha_\pm=0$. Then the projected pairing becomes a singlet pairing of pseudo-spin degrees of freedom, with
\begin{equation}
    \widetilde\Delta({\bf k}) = U^\dagger({\bf k})\Delta({\bf k})U^*({-\bf k}) =-\Delta_0\zeta_y.
\end{equation}
Here, $\zeta_i$ denotes the Pauli matrices for the projected band indices, instead of the physical spin indices. Hence, the projected pairing manifests as a constant \textit{interband} pairing. The complete projected BdG Hamiltonian is given by
\begin{equation}
    H_{proj}({\bf k})=E_0\tau_z \otimes\zeta_z-\Delta_0\tau_x\otimes\zeta_y.
\end{equation}
Notably, the pairing term now commutes with the normal-state Hamiltonian, and therefore, the BdG spectrum is gapless.

\begin{figure*}[t]
    \includegraphics[width=0.95\textwidth]{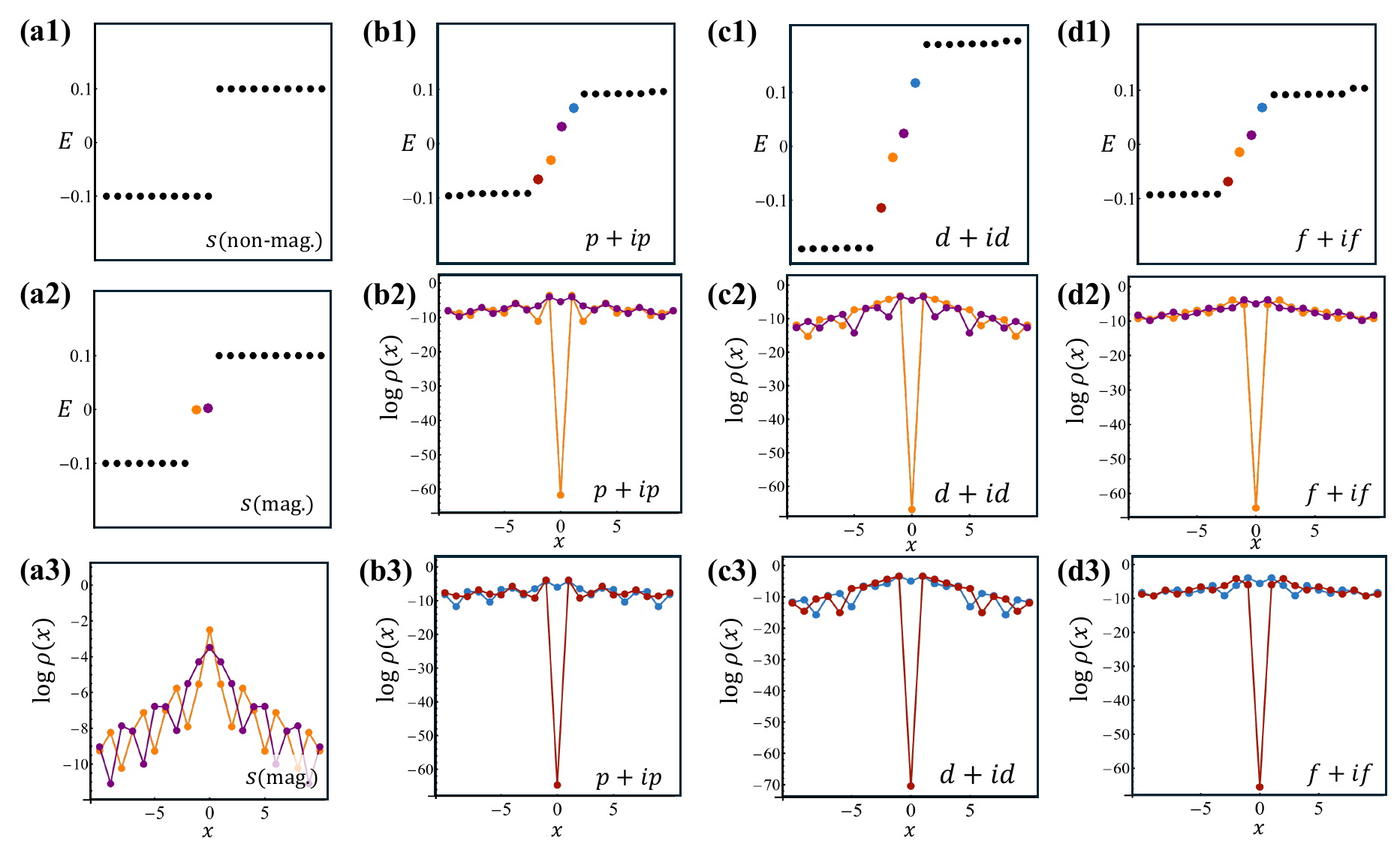}
    \caption{(a1–a2) Energy spectra for an $s$-wave superconductor in the presence of a point-like non-magnetic impurity (a1), and a magnetic impurity (a2). While no in-gap states are found in (a1), two bound states marked by orange and purple dots emerge under magnetic scattering in (a2). (a3) Spatial LDOS plots for the two bound-states in (a2), showing finite weight at the impurity site ($x=0$) and no node–antinode structure. (b1–d1) Energy spectra for chiral $p+ip$, $d+id$, and $f+if$ superconductors, respectively, in the presence of a point-like magnetic impurity. Four non-degenerate in-gap states appear in each case, forming two particle–hole symmetric pairs. (b2–d3) LDOS corresponding to each bound state in (b1–d1), with matched colors. For each pair of opposite-energy states, one component vanishes at the impurity site while the other remains finite, consistent with the expected node–antinode structure.}
    \label{fig:chiral-spdf mag}
\end{figure*}

\section{Magnetic impurity and Robustness of Nodal Condition}

The nodal condition established in Sec.~\ref{sec:analytical proof} applies to both nonmagnetic and magnetic impurities. While the main text focuses on nonmagnetic impurities in chiral superconductors, here we present numerical evidence that magnetic impurities exhibit the same characteristic signatures.

We consider the same square-lattice chiral superconductor model $H_c({\bf k})$ as in the main text, examining the following spin-triplet pairing channels:
(i) an $s$-wave pairing with $d_y({\bf k}) = -i\Delta_0$, corresponding to a phase winding number $l = 0$;
(ii) a chiral $p$-wave pairing with $d_x({\bf k}) = \Delta_0(\sin k_x + i \sin k_y)$ and $l = 1$;
(iii) a chiral $d$-wave pairing with $d_y({\bf k}) = 2i\Delta_0(\cos k_x - \cos k_y - i \sin k_x \sin k_y)$ and $l = 2$;
(iv) a chiral $f$-wave pairing with $d_x({\bf k}) = -\Delta_0 [2 \sin k_x + \cos k_x \sin k_x - 3 \cos k_y \sin k_x - i(2 \sin k_y - 3 \cos k_x \sin k_y + \cos k_y \sin k_y)]$ and $l = 3$. We implement the BdG Hamiltonian on a $51 \times 51$ square lattice with chemical potential $\mu = 0.5$ and pairing amplitude $\Delta_0 = 0.1$.

For the $s$-wave pairing, a nonmagnetic point-like impurity $U_0 \tau_z \delta({\bf r}_0)$ (with $U_0 = 10$) does not induce any in-gap bound states~\cite{balatsky2006RMP}, as shown in Fig.~\ref{fig:chiral-spdf mag} (a1). In contrast, a magnetic point-like impurity $J \tau_z \otimes \sigma_z$ (with $J = 1$) does induce a pair of in-gap states. However, due to the vanishing phase winding $l = 0$, these states do not display a node–antinode structure, consistent with the nodal condition.

Figs.~\ref{fig:chiral-spdf mag} (b1–d1) show the energy spectra for chiral $p$-, $d$-, and $f$-wave superconductors with a magnetic impurity (now with $J = 5$). In all cases, we observe two pairs of in-gap bound states at particle-hole symmetric energies. Each pair exhibits the characteristic node–antinode LDOS pattern at the impurity site, as shown in Figs.~\ref{fig:chiral-spdf mag} (b2–d3), again consistent with the nodal condition in Eq.(5).

\section{Effects of Fermi Surface Topology}

\begin{figure*}[t]
    \includegraphics[width=0.85\textwidth]{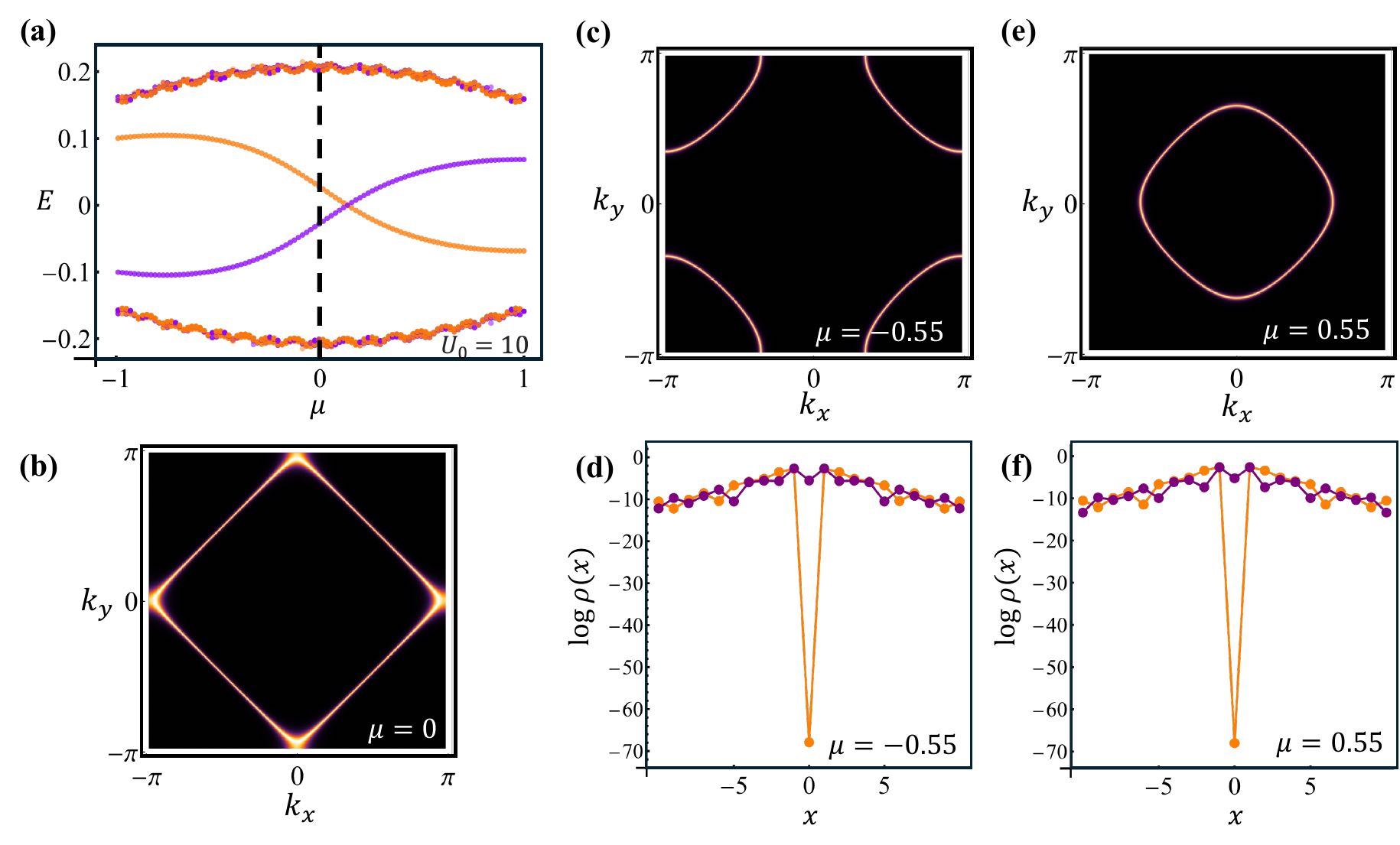}
    \caption{(a) Bound-state spectra of $H_c({\bf k})$ as a function of chemical potential $\mu$ in the presence of a non-magnetic point-like impurity with strength $U_0=10$. The Lifshitz transition occurs at $\mu$ = 0, as shown in the Fermi-surface map in (b). (c) and (e) display the Fermi surfaces for $\mu = -0.55$ and $\mu = +0.55$, respectively. (d) and (f) show the spatial profiles of the impurity-bound states for $\mu = -0.55$ and $\mu = +0.55$, respectively. }
    \label{fig:E-mu Lifshitz transition}
\end{figure*}

In this section, we will numerically prove that the nodal condition of chiral superconductors is resilient against topological changes of electronic Fermi surfaces. In the following, we will investigate two examples considered in the main text: (i) the minimal model $H_c({\bf k})$ and (ii) the BHZ model. 

\subsection{$H_c({\bf k})$}

The normal-state dispersion is given by $\varepsilon({\bf k}) = \cos{k_x} + \cos{k_y} - \mu$, which undergoes a Lifshitz transition at $\mu = 0$. This critical point marks a topological change in the Fermi surface: for $\mu < 0$, the Fermi surface encloses the $M$ points, while for $\mu > 0$, it surrounds the $\Gamma$ point. To visualize the underlying Fermi surface topologies, Fig.~\ref{fig:E-mu Lifshitz transition} (b) displays the Fermi surface at the critical point $\mu = 0$, where electron pockets touch at the $X$ and $Y$ points. For comparison, Figs.~\ref{fig:E-mu Lifshitz transition} (c) and (e) show the Fermi surfaces at $\mu = -0.55$ and $\mu = 0.55$, corresponding to the $M$-centered and $\Gamma$-centered regimes, respectively.

Fig.~\ref{fig:E-mu Lifshitz transition} (a) shows the evolution of the impurity bound-state spectrum as $\mu$ is tuned across this transition. We have included a chiral $d$-wave pairing in the simulation. 
Despite the topological change in the Fermi surface, the impurity-induced bound states maintain the characteristic node–antinode structure, as seen in the LDOS profiles in Figs.~\ref{fig:E-mu Lifshitz transition} (d) and (f). This robustness underscores that the spatial structure of the bound-state wavefunction is dictated primarily by the pairing symmetry, and remains qualitatively intact across the Lifshitz transition. The impurity is located at $x = 0$ in all cases.

\begin{figure*}[t]
    \includegraphics[width=0.6\textwidth]{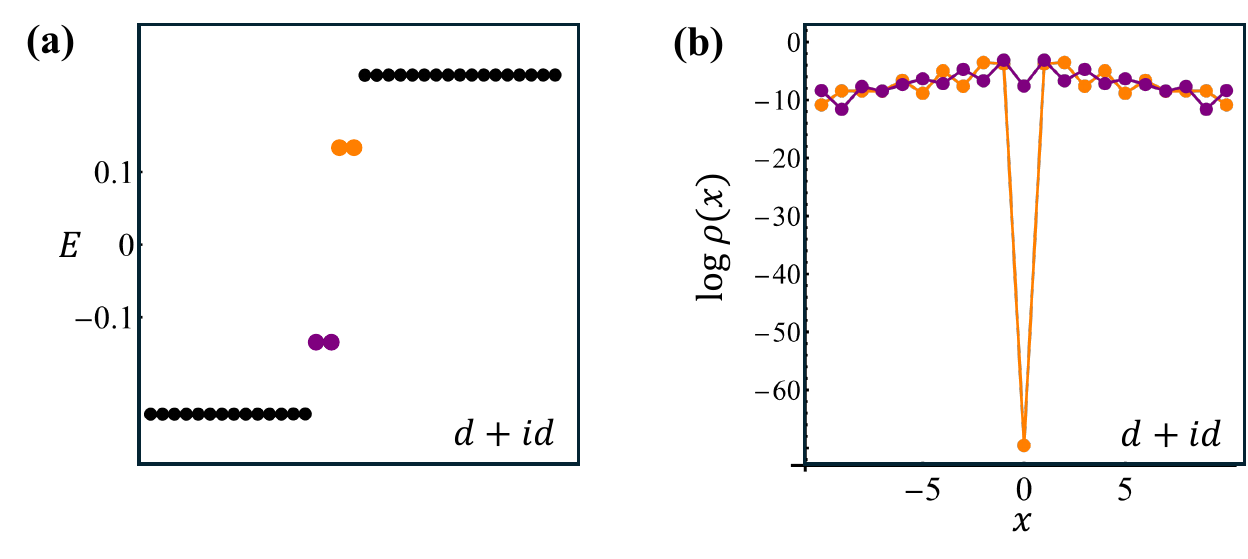}
    \caption{(a) Bound-state spectra of the BHZ model with $\mu=1.5$ in the presence of a non-magnetic point-like impurity and a $d+id$ pairing. (b) displays the LDOS plots for bound states, using the same color scheme as (a).}
    \label{fig:bhz one FS}
\end{figure*}

\subsection{BHZ model}

In the main text, we analyzed the BHZ model at $\mu = 0.4$, where two pairs of spin-degenerate Fermi surfaces are present. Here, we turn to the large-$\mu$ regime, where only a single pair of Fermi surfaces appears. Fig.~\ref{fig:bhz one FS}(a) shows the impurity-induced bound-state spectrum for chiral $d+id$ pairing at $\mu = 1.5$, a regime in which the Fermi surface lies well above the band-inversion region. In this limit, the low-energy physics is effectively single-band in nature, and the impurity response should resemble that of a system with a single Fermi surface. As shown in Fig.~\ref{fig:bhz one FS}(b), the resulting bound states clearly display a node–antinode structure, in full agreement with Eq.(5) of the main text. This again confirms the validity of the nodal condition in capturing the impurity response across different band structures.

\end{document}